\documentclass{aa}  

\usepackage{graphicx}
\usepackage[svgnames]{xcolor}
\usepackage[pdftex,colorlinks,citecolor=DarkBlue]{hyperref}
\usepackage{txfonts}
\usepackage{mathrsfs}
\usepackage{diagbox}
\bibpunct{(}{)}{;}{a}{}{,}
\usepackage[switch]{lineno}
\usepackage{orcidlink}
\begin{document} 

\title{Cosmic-ray electron propagation in the peculiar barred spiral galaxy NGC~2442}


\titlerunning{Cosmic-ray electron propagation in NGC~2442}

\author{Shengtao Wang\orcidlink{0009-0003-9052-1976}\inst{1}
    \and
        Xiaohui Sun\orcidlink{0000-0002-3464-5128}\inst{1} 
    \and
        Volker Heesen\orcidlink{0000-0002-2082-407X}\inst{2}
    \and
        George Heald\orcidlink{0000-0002-2155-6054}\inst{3,4} 
    \and
       Jiangtao Li\orcidlink{0000-0001-6239-3821}\inst{5}
    \and
       Chao-Wei Tsai\orcidlink{0000-0002-9390-9672}\inst{6,7,8}
    \and
        Stefan W.~Duchesne\orcidlink{0000-0002-3846-0315}\inst{4}
    \and
        Andrew J. Battisti\orcidlink{0000-0003-4569-2285}\inst{9,10,11}
    \and 
        Yik Ki Ma\orcidlink{0000-0003-0742-2006}\inst{12}
    \and 
        Amit Seta\orcidlink{0000-0001-9708-0286}\inst{10}
    \and 
        Jun Xu\orcidlink{0000-0003-1778-5580}\inst{6,13}
    \and
        Naomi McClure-Griffiths\orcidlink{0000-0003-2730-957X}\inst{10}
    \and
        Christopher J. Riseley\orcidlink{0000-0002-3369-1085}\inst{14,15}
    \and
       Sam Taziaux\orcidlink{0009-0001-6908-2433}\inst{4,14,15}
    \and 
       Jacco Th. van Loon\orcidlink{0000-0002-1272-3017}\inst{16}
    \and 
        Tayyaba Zafar\orcidlink{0000-0003-3935-7018}\inst{17}
    }

\institute{School of Physics and Astronomy, Yunnan University, Kunming 650500, China\\
\email{wstfch@mail.ynu.edu.cn, xhsun@ynu.edu.cn} 
\and 
Hamburg University, Hamburger Sternwarte, Gojenbergsweg 112, 21029 Hamburg, Germany 
\and
SKA Observatory, SKA-Low Science Operations Centre, 26 Dick Perry Avenue, Kensington, WA 6151, Australia 
\and
CSIRO Space and Astronomy, PO Box 1130, Bentley, WA 6102, Australia 
\and
Purple Mountain Observatory, Chinese Academy of Sciences, 10 Yuanhua Road, Nanjing 210023, China 
\and
National Astronomical Observatories, Chinese Academy of Sciences, 20A Datun Road, Beijing,100101, China 
\and
Institute for Frontiers in Astronomy and Astrophysics, Beijing Normal University, Beijing 102206, China 
\and
School of Astronomy and Space Science, University of Chinese Academy of Sciences, Beijing 100049, China 
\and
International Centre for Radio Astronomy Research, University of Western Australia, 7 Fairway, Crawley, WA 6009, Australia 
\and
Research School of Astronomy and Astrophysics, Australian National University, Canberra, ACT 2611, Australia 
\and
ARC Centre of Excellence for All Sky Astrophysics in 3 Dimensions (ASTRO 3D), Australia 
\and 
Max-Planck-Institut für Radioastronomie, Auf dem Hügel 69, 53121 Bonn, Germany 
\and
National Key Laboratory for Radio Astronomy, Beijing 100101, China 
\and
Ruhr University Bochum, Faculty of Physics and Astronomy, Astronomical Institute (AIRUB), Universitätsstraße 150,
44801 Bochum, Germany 
\and
Ruhr Astroparticle and Plasma Physics Center (RAPP Center), 44780 Bochum, Germany 
\and 
Lennard-Jones Laboratories, Keele University, Keele, ST5 5BG, UK 
\and
School of Mathematical and Physical Sciences, Macquarie University, Sydney, NSW 2109, Australia 
}



\abstract
{Face-on spiral galaxies offer a favorable geometry for studying magnetic-field structures and their impact on cosmic-ray (CR) propagation, as projection effects and structural overlap are significantly reduced.}
{We aim to investigate the transport of cosmic-ray electrons (CREs) in the nearby face-on spiral galaxy NGC~2442 and to assess how the galactic environment and magnetic-field structure influence their propagation.}
{We combined radio continuum (RC) observations from Australian SKA Pathfinder (ASKAP) at 943~MHz, MeerKAT at 1.28 and 1.7~GHz, and ATCA at 5~GHz with optical H$\alpha$ and infrared data to characterize the galactic environment. We then compared these observations with 2D CRE transport simulations to evaluate the role of the magnetic field and the surrounding environment in shaping CRE propagation.}
{NGC~2442 exhibits an integrated radio continuum spectrum that is steeper than those of most nearby galaxies, with spectral indices of $\alpha=-0.96\pm0.04$ for the total emission and $\alpha_{\rm nt}=-1.21\pm0.04$ for the synchrotron emission over 408 MHz--5 GHz. The spectrum shows a break near 1~GHz, indicating substantial radiative aging of the CRE population. Under the equipartition assumption, we derived a mean magnetic-field strength of $10.8\,\mu{\rm G}$. The RC--star formation rate (SFR) smoothing analysis indicates effective CRE propagation lengths of $\sim 0.65$--$0.89$~kpc at 943--1700~MHz and $\sim 0.44$~kpc at 5~GHz, corresponding to effective diffusion coefficients of the order of $10^{28}\,{\rm cm^2\,s^{-1}}$. We identify a steep-spectrum synchrotron ``island'' in the southeastern region, with an average synchrotron spectral index of $\alpha \sim  -1.09$ and no clear H$\alpha$, infrared, far-ultraviolet (FUV) or near-ultraviolet (NUV) counterpart, indicating that the CREs are unlikely to be injected in situ. Our 2D simulations using CRPropa show that anisotropic diffusion along ordered magnetic fields allows CREs to reach the island region much more efficiently than isotropic diffusion.}
{NGC~2442 provides evidence that the combined effects of environmental disturbances and ordered magnetic fields can strongly regulate CRE propagation in disturbed spiral galaxies.}

\maketitle
\section{Introduction}
\label{sec:intro}

Cosmic rays (CRs) are an important component of the interstellar medium (ISM), contributing significantly to its total pressure and energy budget and thereby influencing the dynamical evolution of galaxies~\citep{zweibel2017,hopkins2020,ruszkowski2023}. In particular, cosmic-ray electrons (CREs) are powerful tracers of CR transport because they emit synchrotron radiation in galactic magnetic fields and lose energy on observable timescales through synchrotron and inverse-Compton (IC) losses. Radio continuum (RC) observations, therefore, provide a direct way to study the propagation of CREs and the interplay between star formation, magnetic fields, and the surrounding ISM~\citep{lisenfeld2000,heesen2021}. At frequencies from a few hundred MHz to about 1~GHz, RC emission in normal star-forming galaxies is largely dominated by non-thermal synchrotron radiation. At GHz frequencies, the combination of total intensity, thermal fraction, and spectral index information makes it possible to constrain CRE aging and transport~\citep{heesen2019}. 

Understanding how CREs propagate through galactic disks and halos is essential for incorporating CR feedback into galaxy-evolution models. In principle, CR transport may occur through diffusion, advection, and streaming, and the relative importance of these processes depends strongly on the magnetic-field structure and the physical conditions of the ISM. A proper treatment of these transport mechanisms is therefore essential for incorporating CRs into models of galaxy evolution \citep{uhlig2012,wiener2017,hopkins2020}. 

Edge-on galaxies are particularly useful for studying vertical CR transport and halo outflows, and have provided measurements of diffusion coefficients and advection speeds in radio halos~\citep{heesen2016,heesen2019,schmidt2019,stein_2019a}. Face-on galaxies, on the other hand, offer a complementary perspective because projection effects are minimized and spiral arms can be separated more cleanly. In such systems, CRE transport has commonly been studied either by modeling the radial distribution of the radio spectral index in terms of CR diffusion or by comparing the RC emission with star-formation tracers using smoothing analyses~\citep{bicay1990,murphy2008,mulcahy2016,heesen2014,heesen2023,dorner2023}. In the latter approach, the star-formation map is convolved with a suitable kernel to minimize the differences between the two maps or to linearize their spatial correlation, thereby estimating the characteristic CRE propagation length. However, these methods have several limitations. They generally neglect halo emission projected along the line of sight and treat CRE transport in a simplified, effectively two-dimensional way. In addition,~\citet{mulcahy2016} showed that CRE escape must be included in order to reproduce the observed RC spectrum. While such diffusion-based approaches have provided valuable insight into CRE transport in face-on systems, a more realistic description requires a 3D treatment that accounts for the magnetic-field geometry and the possibility of anisotropic transport. This is particularly important in disturbed galaxies, where environmental effects may reshape both the magnetic field structure and spatial distribution and transport of CREs.

Environmental effects are known to play an important role in the evolution of galaxies in both groups and clusters. In clusters, galaxies are affected by a variety of processes that can modify their star formation activity, leading to H\,\textsc{i} deficiency, truncation of the H$\alpha$-emitting gas disk, morphological transformation, and strong dynamical evolution~\citep{boselli2006}. The interstellar medium (ISM) of disk galaxies can be altered by ram-pressure stripping and gravitational interactions, which redistribute both the stellar and gaseous components of the disk~\citep{kenney2004,vollmer2001}. Similar environmental processes may also influence galaxies in group environments.

\begin{figure}	
    \centering
    \includegraphics[width=0.98\columnwidth]{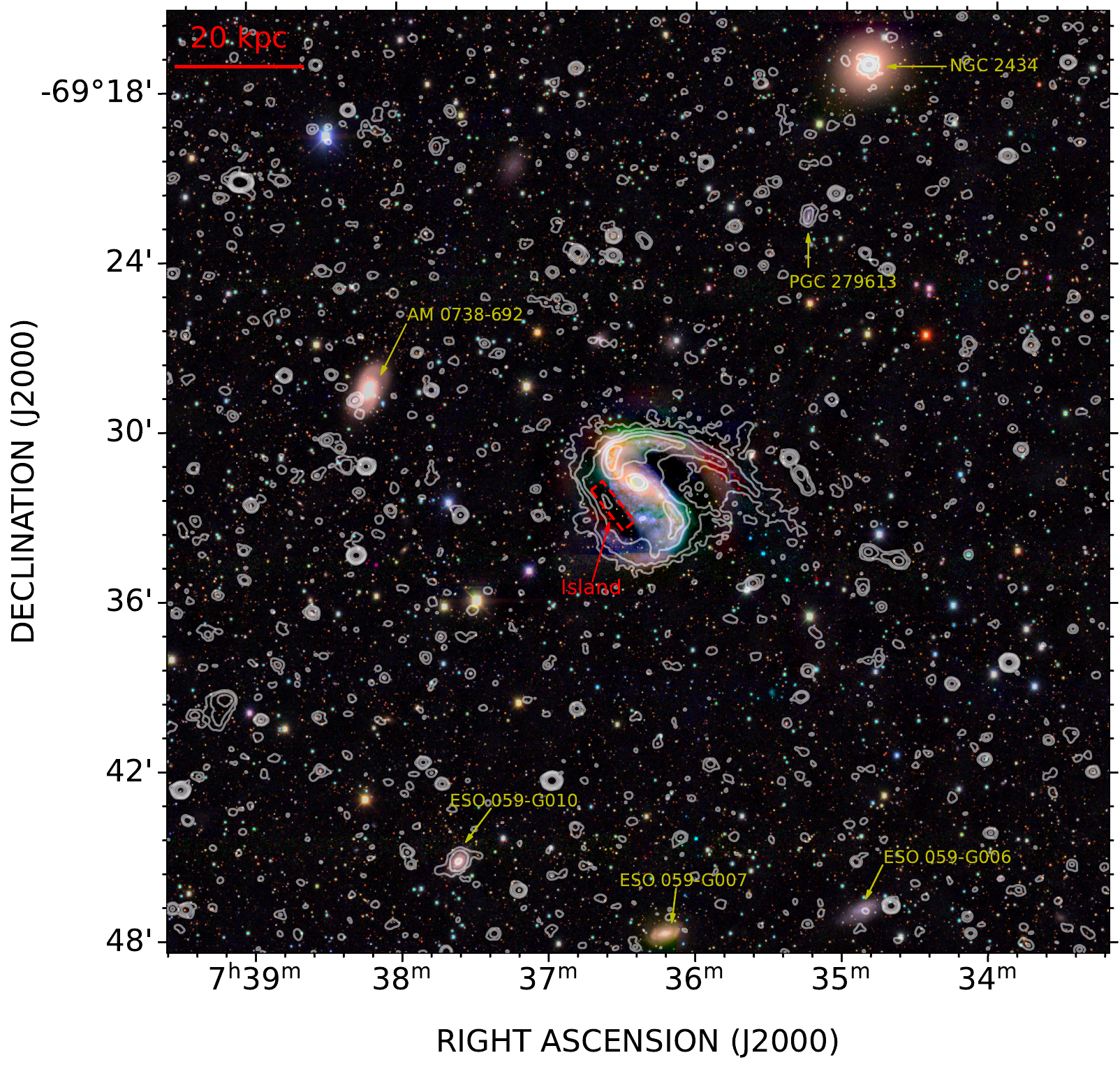}
    \caption{ASKAP total-intensity contours at 943~MHz are overlaid on a three-color SDSS optical image constructed from the i, r, and g filters. Other galaxies in the NGC~2442 group are indicated. The red dashed rectangle marks  the ``island'' region.}
\label{fig:optical_radio}
\end{figure}

NGC~2442 is a nearby peculiar barred spiral galaxy with an asymmetric morphology and resides in a disturbed group environment~\citep{ryder2001,saponara2025}. Its northern spiral arm is compressed, elongated, and extends to projected distances exceeding 20~kpc from the main body, accompanied by a prominent dust lane, whereas the southern arm is broader and more open, reaching only $\sim13$~kpc along its spine and exhibiting a more chaotic dust-lane structure (\citealt{mihos1997}; \citealt{harnett2004}; see also Fig.~\ref{fig:optical_radio}). Previous studies have suggested that this disturbed appearance is related to tidal interaction and/or ram pressure effects in the group environment~\citep{mihos1997,ryder2001}. If NGC~2442 experienced a recent tidal encounter, the existence of HIPASS J0731-69 suggests that a more massive galaxy to the northwest may have been responsible, such as the elliptical galaxy NGC~2434 or perhaps even the spiral plus Magellanic irregular pairing of NGC~2397 and NGC~2397A~\citep{ryder2001}. The velocity field also shows evidence of non-circular motions, especially around the northern arm~\citep{bajaja1999}. The nuclear region can be resolved into a circumnuclear ring with $\sim0.8$~kpc radius in the Spitzer Infrared Array Camera (IRAC) images~\citep{pancoast2010}, which was also suggested by~\citet{mihos1997} based on their H$\alpha$ map of the galaxy, and the nucleus has a LINER spectrum~\citep{bajaja1999}.

Previous radio studies showed that NGC~2442 has an integrated spectral index of $\alpha=-0.92\pm0.08$ over the frequency range 408~MHz--5~GHz, which is unusually steep for a spiral galaxy~\citep{harnett1984}. The Molonglo Observatory Synthesis Telescope (MOST) images further revealed a large-scale non-thermal feature to the east of the galaxy~\citep{harnett1984}, visible in Fig.~\ref{fig:optical_radio}, with no obvious optical or H$\alpha$ counterpart. Its extremely high fractional polarization of $\sim52\%$ at 6~cm makes it particularly prominent in polarized intensity, and this region has become known as the ``island''~\citep{harnett2004}. Together, these characteristics make NGC~2442 an excellent laboratory for investigating how a disturbed group environment and ordered magnetic fields shape CRE propagation.

In this work, we investigate the transport of CREs in NGC~2442 using new and archival multi-frequency RC data, including Australian SKA Pathfinder (ASKAP) observations at 943~MHz, MeerKAT observations at 1.28 and 1.7~GHz, and ATCA observations at 5~GHz, together with H$\alpha$ and infrared data tracing the star-forming environment. We use these data to derive the thermal and non-thermal RC components, spatially resolved spectral index distributions, magnetic-field strengths under the equipartition assumption, and characteristic CRE propagation lengths from the RC--star formation rate (SFR) relation. We further compare the observational results with 2D CRE transport simulations in order to explore the role of anisotropic magnetic fields and the disturbed environment in shaping the CRE distribution. We aim to address how the peculiar structure of NGC~2442, including its distorted spiral arms and the southeastern synchrotron ``island'', is linked to the transport and aging of CREs.

This paper is organized as follows. The data acquisition and reprocessing are described in Sect.~\ref{sec:ob_data}. The results are presented in Sect.~\ref{sec:results}. The discussion is given in Sect.~\ref{sec:discussions}, and the conclusions are summarized in Sect.~\ref{sec:conclu}.

\begin{table}
	\centering
	\caption{Parameters of NGC~2442.}
	\label{tab:parameters_table}
	\begin{tabular}{lccr} 
		\hline
		\hline
            RA               &$\rm {07^h36^m23\fs8}$ \\
            Dec              & $-69\degr31\arcmin51\farcs0$ \\
            $D$ (Mpc)\tablefootmark{a}               & $15.5$ \\
            $i$ (from face-on)\tablefootmark{b}      & $24\degr$\\
            PA\tablefootmark{b}                      & $40\degr$\\
		Morphological type\tablefootmark{c}      & $\rm SBbc (rs)$ \\
            $D_{25}\,(\arcmin)$\tablefootmark{c}   &  5.5 \\
            $\mathrm{log}\,L_{\rm TIR}\,(L_{\odot})$\tablefootmark{d}      & $10.3$ \\
            ${V_{\rm sys}\,\rm(km\,s^{-1})}$\tablefootmark{a}  &$1449$     \\
            ${V_{\rm rot}\,\rm(km\,s^{-1})}$\tablefootmark{e}  &$225$     \\
            $\mathrm{log}\,M_{\star}$ ($\rm M_{\odot}$)\tablefootmark{f} & 10.6 \\
            $\mathrm{log}\,M_{\rm HI}$ ($\rm M_{\odot}$)\tablefootmark{a} & 9 \\
            $\rm SFR$ ($\rm M_{\odot}\,yr^{-1}$)\tablefootmark{g} & $6-7$ \\
            $\Sigma_{\rm SFR}$ ($\rm M_{\odot}\,yr^{-1}\, kpc^{-2}$)\tablefootmark{g} & $0.062$\\
            $B\,(\rm \mu G)$\tablefootmark{h}         & $10.8$ \\
            $\alpha$\tablefootmark{h}                & $-1$~--~$-0.8$ \\
		\hline
	\end{tabular}
     \tablefoot{
    \textnormal{\tablefoottext{a}{Distance to the galaxy~\citep{ryder2001}.}\tablefoottext{b}{Inclination angle and position angle of the galaxy~\citep{bajaja1999}.} \tablefoottext{c}
    {Optical diameter in the $B$ band~\citep{harnett2004}.} \tablefoottext{d}{The total infrared luminosity $L_{\rm TIR}=L(8-1000\,\rm \mu m)$~\citep{sanders2003}.}\tablefoottext{e}{The rotation velocity~\citep{mihos1997}.} \tablefoottext{f}{The stellar mass was estimated using WISE W1 and W2 photometry~\citep{cluver2014}.}\tablefoottext{g}{The star formation rate~\citep{pancoast2010}.}\tablefoottext{h}{The average magnetic field strength and radio spectral index are derived from this work.}}
      }
\end{table}

\section{Data acquisition and reprocessing}
\label{sec:ob_data}
\subsection{ASKAP data}

\begin{figure}	
    \centering
    \includegraphics[width=0.98\columnwidth]{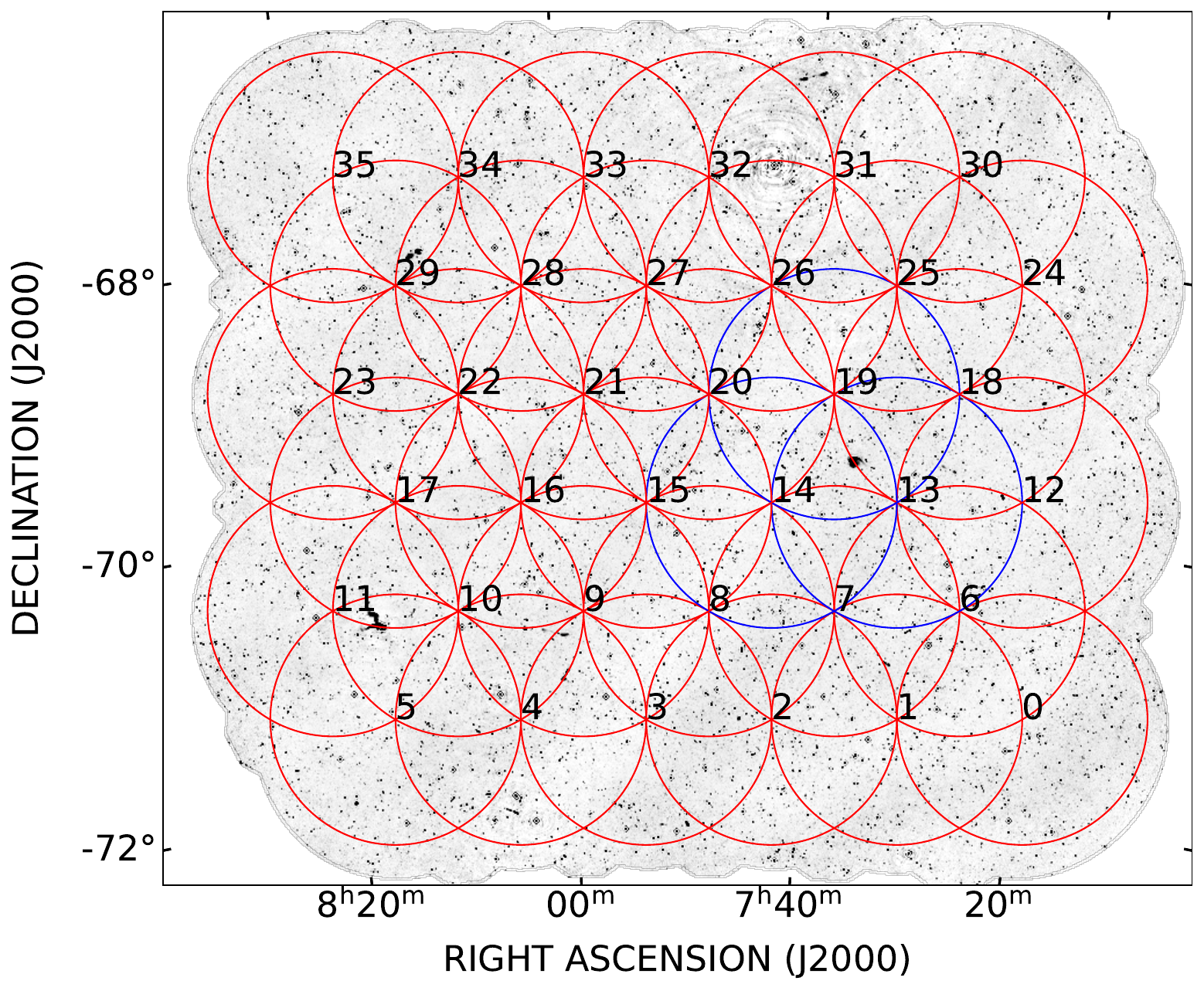}
    \caption{Layout of the 36 ASKAP beams with circles in the ``closepack36'' footprint configuration, overlaid on the $I$ image from CASDA. The radius of the circles is $0\fdg9$, which is approximately the primary beam width. The blue circles indicate the beams that cover NGC~2442.}
\label{fig:36beam_cover}
\end{figure}

 NGC~2442 lies within a field observed as part of the ASKAP Evolutionary Map of the Universe (EMU) survey~\citep{hopkins2025} (Project ID: AS201; Observation ID: SB59742), with a total on-source integration time of 10~h. The central frequency is 943~MHz, and the total bandwidth of 288~MHz is divided into 1-MHz channels. Each ASKAP antenna is equipped with a phased array feed that forms 36 dual-polarization beams~\citep{hotan2021}. The footprints of the beams are in the ``closepack36'' configuration as shown in Fig.~\ref{fig:36beam_cover}. NGC~2442 was fully covered by beams 13, 14, and 19, outlined by blue circles in Fig.~\ref{fig:36beam_cover}. 

The calibrated visibilities for each individual beam and the total intensity ($I$) images for the whole field, processed with {\small ASKAPsoft}\footnote{ASKAPsoft is the suite of processing software developed by the ASKAP computing team to process ASKAP observations.}, are public and available from the CSIRO ASKAP Science Data Archive \textit \rm {(CASDA)}\footnote{\url{https://data.csiro.au/domain/casdaObservation}}. We reprocessed the visibility data for beams 13, 14, and 19 from CASDA using the pipeline SASKAP\footnote{\url{https://gitlab.com/Sunmish/saskap/-/tree/petrichor}} to improve calibration and imaging. 

For calibration, we performed two rounds of phase-only (p) and one round of amplitude-phase (ap) self-calibration. The solution interval was progressively reduced during each round, decreasing from 300~s initially to 60~s in the final round. 

For imaging, the fast generic widefield imager \textit{\rm {\small \textsc{WSclean}}}\footnote{\url{https://wsclean.readthedocs.io/en/latest/changelogs/v3.2.html}}~\citep{offringa2014,offringa2017} was used to generate a multi-frequency and multiscale synthesis image. Briggs weighting~\citep{briggs1995} with ${\rm robust = 0.25}$ was employed to enhance sensitivity toward faint, extended emission. Wgridder was employed to improve the accuracy of wide-field image reconstruction and to efficiently account for non-coplanar baseline effects~\citep{ye2022,arras2021}.

Primary beam correction is required to determine flux densities and combine images of different beams. Following the method of~\citet{duchesne2024}, we built a robust model of the primary beam. We cross-matched the point sources with the Rapid ASKAP Continuum Survey low-frequency \textit \rm {(RACS-Low)}\footnote{\url{https://data.csiro.au/collection/csiro\%3A52217v3}}~\citep{mcconnell2020,hale2021}, and obtained the ratio of flux densities as a function of the position relative to the beam center. A two-dimensional elliptical Gaussian was fit to the data to obtain the primary beam model, which was used to correct the images. Finally, we performed a weighted linear combination of the three images after primary-beam correction to improve the signal-to-noise ratio.

\subsection{MeerKAT data}
MeerKAT is a radio interferometer located in the Karoo desert of South Africa, consisting of 64 dishes with baselines out to 8~km~\citep{jonas2009,jonas2016,mauch2020}. The parabolic 13.5~m diameter antennas have offset Gregorian receivers. Of the 64 antennas, 48 are located in the inner core within a 1~km radius, providing the shortest baseline of 29~m. 

The MeerKAT 1.28~GHz observations were taken from the MeerKAT atlas of southern Revised Bright Galaxy Sample (RBGS) sources~\citep{condon2021}. These observations consist of five 3~min snapshots, corresponding to a total on-source integration time of 15~min per galaxy. The resulting images have an angular resolution of approximately $7\farcs5$ and an rms noise level of $20~\mu{\rm Jy~beam^{-1}}$.

We also used the MeerKAT 1.7~GHz RC image from~\citet{saponara2025}. Their data cover 856--1712~MHz and are divided into ten frequency channels. We directly adopted their fully calibrated and imaged FITS products, but only used the 1.7~GHz image here because it is less affected by imaging artifacts than the images at other frequencies. The native synthesized beam of the 1.7~GHz image is $8\farcs72\times 6\farcs43$, with a position angle of $0\degr$. For comparison with the other radio data, both MeerKAT images were convolved to a common angular resolution of $15\arcsec$.

\subsection{ATCA data}

We used archival ATCA 5-GHz RC observations of NGC~2442 from~\citet{harnett2004}. The observations were obtained at a central frequency of 5170~MHz with a total bandwidth of 128~MHz, using the 750~m and 375~m array configurations. The data were taken over five observing runs between 1996 and 2000, with a total on-source integration time of about 37~h. The original total-intensity and polarized-intensity maps were produced with an angular resolution of $10\arcsec$ and an rms noise of about $25~\mu{\rm Jy~beam^{-1}}$. In this work, we used the total-intensity image and convolved it to the common resolution of $15\arcsec$ for comparison with the ASKAP and MeerKAT data.

\subsection{Largest angular scales and flux recovery}

The shortest baseline of ASKAP is 22~m, corresponding to the largest detectable angular scale of about $49\farcm7$ at 943~MHz. For MeerKAT, the shortest baseline is approximately 29~m, providing sensitivity to diffuse emission on angular scales of up to $15$--$25\arcmin$ across the L band. The shortest baseline of the ATCA 375~m configuration is approximately 31~m, corresponding to a maximum recoverable angular scale of about $6\farcm6$ at 5~GHz. Since the
optical angular extent of NGC~2442 is approximately $5\farcm5$, no significant loss of large-scale flux is expected in these observations.

\subsection{Auxiliary data}

Additional RC data were collected from the literature to construct the integrated spectrum of NGC~2442. These include low-frequency measurements from 76 to 227~MHz obtained with the Murchison Widefield Array (MWA) as part of the GaLactic and
Extragalactic All-sky MWA survey (GLEAM)~\citep{Hurley-Walker+17}, a
408~MHz measurement from the cross-type radio telescope of the Molonglo Radio Observatory \citep{cameron1971}, and a 2700~MHz measurement from Parkes~\citep{wright1974}. The integrated flux densities from these surveys were used to constrain the radio spectral index and are summarized in Table~\ref{tab:int_flux}.

We complemented our RC data with optical integral-field spectroscopy from the \textit \rm{TYPHOON}\footnote{\url{https://typhoon.datacentral.org.au/}} project~\citep{grasha2022}, GALEX far-ultraviolet (FUV) and near-ultraviolet (NUV) imaging, Spitzer/MIPS 24~$\mu$m imaging, and Spitzer/IRAC 3.6, 4.5, and 8~$\mu$m imaging. The TYPHOON H$\alpha$ map was used to trace unobscured star formation and to estimate the thermal free--free emission. The TYPHOON data have a spatial sampling of $1\farcs65$ pixel$^{-1}$, while the effective angular resolution is set by the seeing of the observations. The GALEX FUV and NUV images were taken from the $z=0$ Multiwavelength Galaxy Synthesis (z0MGS) GALEX--WISE atlas~\citep{leroy2019}, which provides uniformly processed UV and infrared images of nearby galaxies based on GALEX and WISE observations, with matched astrometry, background subtraction, and products convolved to common angular resolutions. In this work, we used the $7\farcs5$ resolution products. The Spitzer/MIPS 24~$\mu$m image has an angular resolution of approximately $6\arcsec$. The Spitzer/IRAC images were used to construct a three-color infrared image for visualization and comparison with the radio and UV morphology. The maps used in the thermal-emission and RC--SFR analyses were regridded to the same astrometric frame as the radio images and convolved to a common angular resolution of $15\arcsec$.

\begin{figure*}[!htbp]
    \centering
    \includegraphics[width=0.98\textwidth]{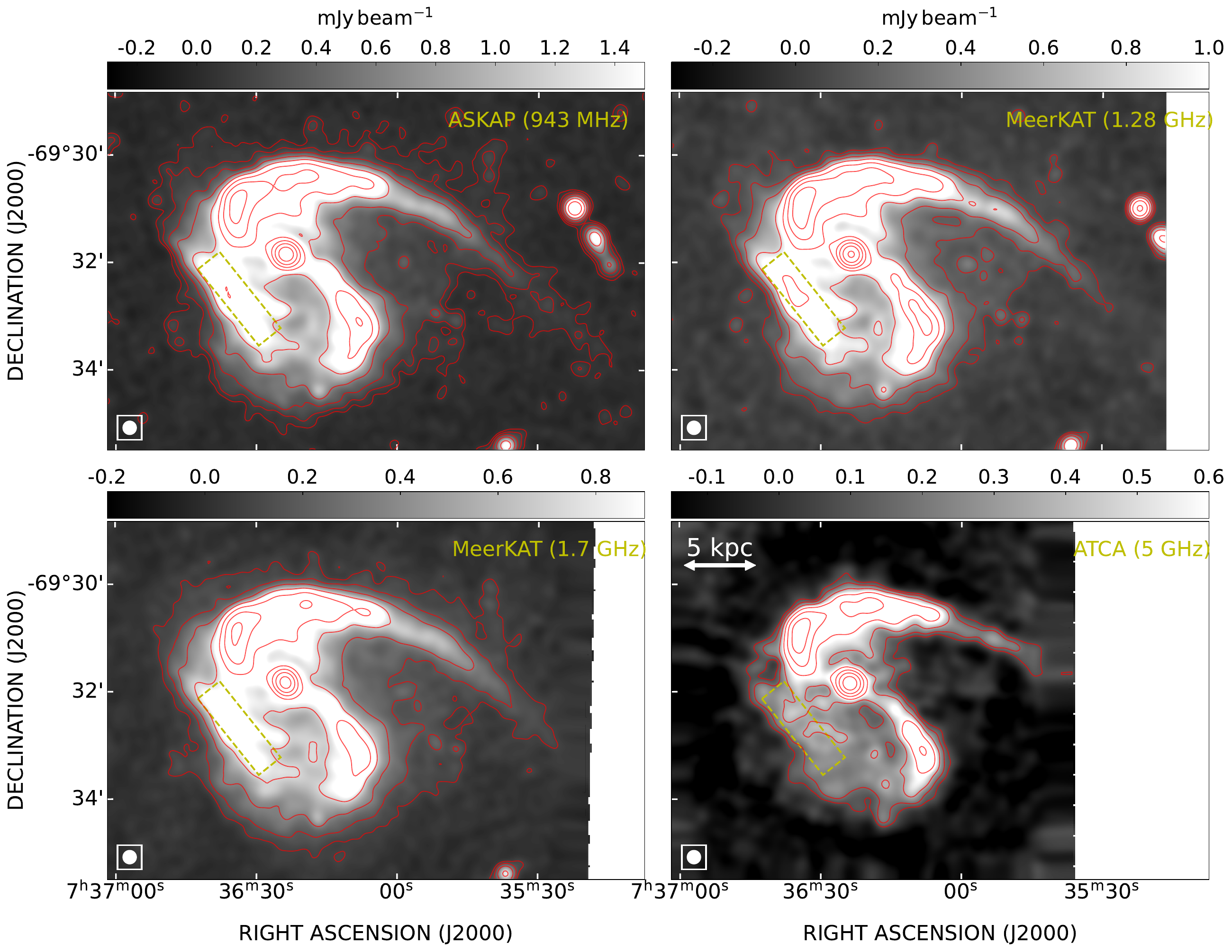}
    \caption{Total-intensity images of NGC~2442 from ASKAP at 943~MHz (top left), MeerKAT at 1.28~GHz (top right), MeerKAT at 1.7~GHz (bottom left), and ATCA at 5~GHz (bottom right), with rms noise levels of 25, 34, 30, and $30\,\mu{\rm Jy\,beam^{-1}}$, respectively. All images were convolved to a common angular resolution of $15\arcsec$. Contours are drawn at levels of $3\sigma \times 2^{n}$ ($n = 0, 1, 2, \ldots$). The yellow rectangular box outlines the island region.} 
\label{fig:total_intensity}
\end{figure*}

\section{Results}
\label{sec:results}

\subsection{Total intensity images and integrated flux densities}

In Fig.~\ref{fig:total_intensity}, we present total intensity images convolved to a common angular resolution of $15\arcsec$. From the top left to the bottom right, the panels show images from ASKAP at 943~MHz, MeerKAT at 1.28~GHz, MeerKAT at 1.7~GHz, and ATCA at 5~GHz, with rms noise levels of 25, 34, 30, and 30~$\mu$Jy~beam$^{-1}$, respectively. The distorted morphology of NGC~2442 is visible in all maps. The northern side exhibits compressed and elongated spiral arms with extended tails, while a synchrotron island is clearly visible in the southeastern region (yellow rectangle), with a sharp truncation on its eastern side. The possible environmental origin of these morphological features is discussed in Sect.~\ref{sec:discussions}. In addition, the bright AGN-dominated core is detected.

We derived the integrated flux densities at 943~MHz, 1.28~GHz, 1.7~GHz, and 5~GHz from the total intensity images, assuming relative calibration uncertainties of 5\% for the ASKAP, MeerKAT, ATCA measurements and 10\% for the MWA measurements. All flux densities $S_\nu$ at frequencies $\nu$ are listed in Table~\ref{tab:int_flux}.

\begin{table}[!htbp]
\centering
\caption{Integrated flux density and thermal fractions of NGC~2442.\label{tab:int_flux}}
\begin{tabular}{ccccccc}
\hline\hline
Telescope& ${\rm \nu}$ & $S_{\rm \nu}$ & $f_{th}$ & Ref \\
&(MHz) & (mJy) & (\%) &  \\
\hline
MWA (GLEAM)  & 76  & $2020\pm 202$ &2.9& \tablefootmark{a} \\
MWA (GLEAM)  & 84  & $1840\pm 184$ &3.2& \tablefootmark{a} \\
MWA (GLEAM)  & 92  & $1878\pm 188$ &3.1& \tablefootmark{a} \\
MWA (GLEAM)  & 99  & $1793\pm 179$ &3.2& \tablefootmark{a} \\
MWA (GLEAM)  & 107 & $1631\pm 163$ &3.5& \tablefootmark{a} \\
MWA (GLEAM)  & 115 & $1565\pm 157$ &3.6& \tablefootmark{a} \\
MWA (GLEAM)  & 122 & $1475\pm 148$ &3.8& \tablefootmark{a} \\
MWA (GLEAM)  & 130 & $1472\pm 147$ &3.8& \tablefootmark{a} \\
MWA (GLEAM)  & 143 & $1275\pm 128$ &4.3& \tablefootmark{a} \\
MWA (GLEAM)  & 151 & $1282\pm 128$ &4.3& \tablefootmark{a} \\
MWA (GLEAM)  & 158 & $1215\pm 122$ &4.5& \tablefootmark{a} \\
MWA (GLEAM)  & 166 & $1200\pm 120$ &4.6& \tablefootmark{a} \\
MWA (GLEAM)  & 174 & $1131\pm 113$ &4.8& \tablefootmark{a} \\
MWA (GLEAM)  & 181 & $1136\pm 114$ &4.8& \tablefootmark{a} \\
MWA (GLEAM)  & 189 & $1044\pm 104$ &5.2& \tablefootmark{a} \\
MWA (GLEAM)  & 197 & $1015\pm 102$ &5.3& \tablefootmark{a} \\
MWA (GLEAM)  & 204 & $1190\pm 119$ &4.5& \tablefootmark{a} \\
MWA (GLEAM)  & 212 & $1071\pm 107$ &5.0& \tablefootmark{a} \\
MWA (GLEAM)  & 220 & $1079\pm 108$ &5.0& \tablefootmark{a} \\
MWA (GLEAM)  & 227 & $1036\pm 104$ &5.1& \tablefootmark{a} \\
Molonglo Cross& 408& $900\pm 100$  &5.6& \tablefootmark{b} \\
ASKAP (RACS) & 888 & $472\pm 24$   &9.8& \tablefootmark{c} \\
ASKAP (EMU)  & 943 & $445\pm 23$   &10.4& \tablefootmark{c} \\
MeerKAT      & 1280& $336\pm 17$   &13.3& \tablefootmark{c} \\
MeerKAT      & 1714& $252\pm 13$   &17.3& \tablefootmark{c} \\
Parkes       & 2700& $158\pm 20$   &26.3& \tablefootmark{d} \\
ATCA         & 5000& $88 \pm 5$    &44.4& \tablefootmark{c} \\
\hline
\end{tabular}
\tablefoot{
    \textnormal{$S_\nu$ is the integrated total flux density at frequency $\nu$. \tablefoottext{a}{From GLEAM source catalog~\citep{Hurley-Walker+17}.} \tablefoottext{b}
{The cross-type radio telescope of the Molonglo Radio Observatory~\citep{cameron1971}.}\tablefoottext{c}{From this work.}\tablefoottext{d}{\citet{wright1974}.}}
      }
\end{table}

\subsection{Thermal emission}
\label{sec:therm_em}
Thermal emission in the radio band is dominated by free–free emission that can be estimated using the ``mixture method'' by combining the $\rm H\alpha$ and $\rm 24\,\mu m$ data. Star formation can be traced by these two types of radiation in both obscured and unobscured regions \citep{kennicutt2007}. The corrected H$\alpha$ flux $F_{\rm H\alpha}$ can be derived from the observed flux $F_{\rm H\alpha,\,obs}$ and intensity at 24~$\mu$m $I_{\rm 24\,\mu m}$ as~\citep{kennicutt2009}:
\begin{equation}
    F_{\rm H\alpha} = F_{\rm H\alpha,obs} + 0.042 \cdot \nu_{24\,\rm \mu m}\, I_{24\,\rm{\mu m}}.
    \label{Halpha+24microns}
\end{equation}

\begin{figure*}[!htbp]	
    \centering    \includegraphics[width=\linewidth]{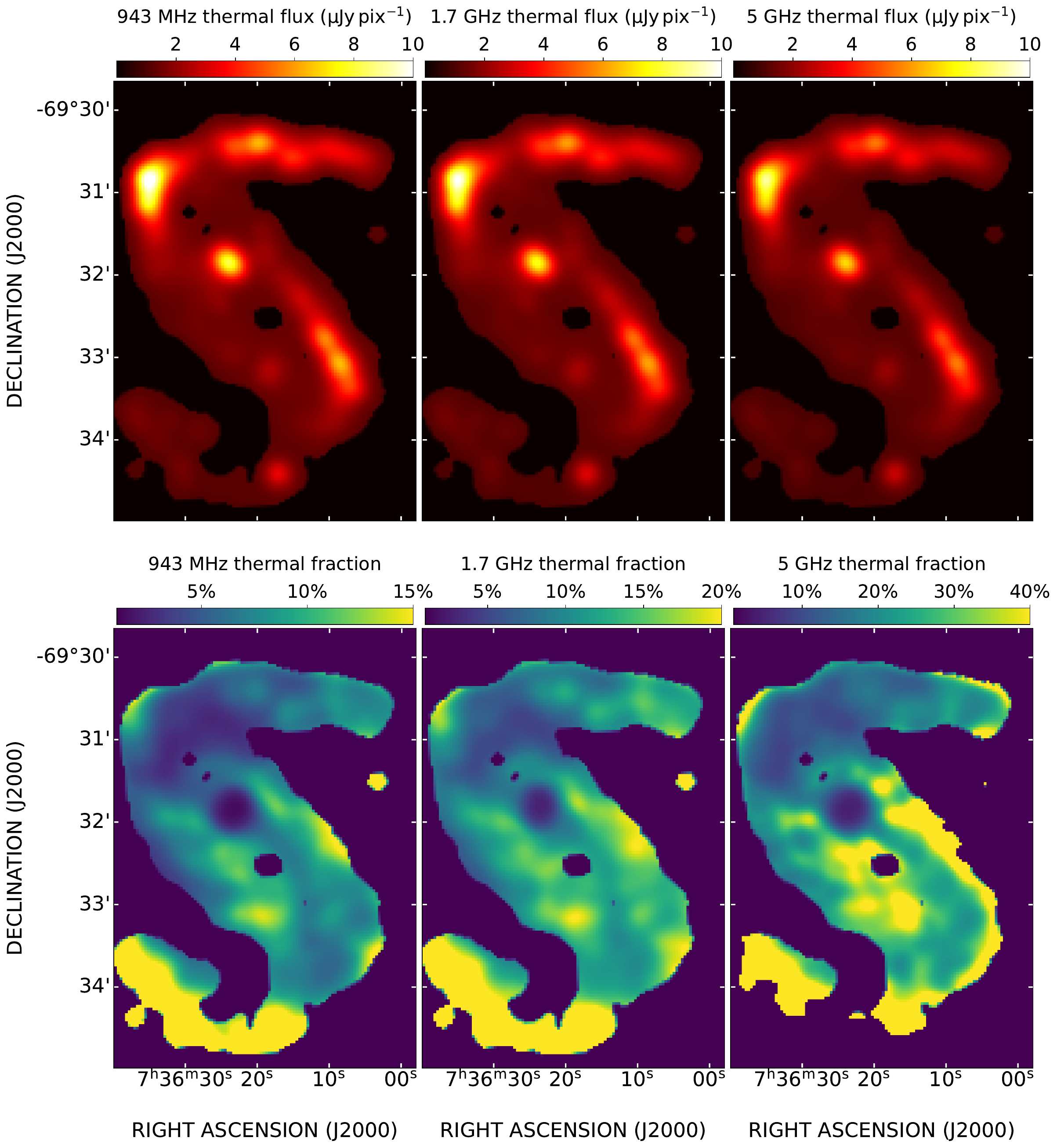}
    \caption{Thermal emission distributions of NGC~2442 at 943~MHz, 1.7~GHz, and 5~GHz are shown in the top panels, and the corresponding thermal fraction maps are presented in the bottom panels.} 
\label{fig:Thermal_emission}
\end{figure*}

The thermal contribution to the RC emission at a given frequency $\nu$ can then be calculated using~\citep{deeg1997}
\begin{align}
\frac{S_{\mathrm{th}}(\nu)}{\mathrm{erg\,cm^{-2}\,s^{-1}\,Hz^{-1}}}
&= 1.14 \times 10^{-14}
   \left( \frac{\nu}{\mathrm{GHz}} \right)^{-0.1} \nonumber \\
&\quad \times \left( \frac{T_e}{10^4\,\mathrm{K}} \right)^{0.34}
   \left( \frac{F_{\mathrm{H}\alpha}}{\mathrm{erg\,cm^{-2}\,s^{-1}}} \right).
\label{radioequation}
\end{align}
The electron temperature was assumed to be $T_e=10^4$~K, a standard value for H\,{\sc ii} regions and the warm ionized medium \citep[e.g.][]{osterbrock2006,Tabatabaei2007}.
We derived the thermal emission distribution at 943~MHz, 1.7~GHz, and 5~GHz, together with the corresponding thermal fraction, as shown in Fig.~\ref{fig:Thermal_emission}. The figure shows that the spiral arms and the central region exhibit stronger thermal emission than the rest of the galaxy. In the spiral arms, this is consistent with enhanced star formation activity, while in the central region the thermal emission may also include contributions from the circumnuclear region and nuclear activity. In contrast, the thermal fraction is low along the spiral arms compared to the inner and outer disk, and reaches a minimum in the nuclear region. This suggests that synchrotron emission dominates the RC budget in much of the disk, especially along the spiral arms. However, this result should be interpreted with caution in dusty regions. Although the mixture H$\alpha$+24~$\mu$m method is widely used to correct for dust-obscured star formation \citep[e.g.][]{calzetti2007,kennicutt2009}, local uncertainties may remain along prominent dust lanes, where strong
H$\alpha$ attenuation and possible mid-infrared radiative-transfer effects could lead to an underestimate of the free--free contribution
\citep{vargas2018}.

Using the integrated flux densities, we derived the thermal fractions of NGC~2442 at each frequency, which are summarized in Table~\ref{tab:int_flux}. As our analysis is primarily concerned with the non-thermal synchrotron emission, the thermal emission was removed from the total RC intensity in the following analysis.

\subsection{Spectral index}
\label{subsec:spex}
\subsubsection{Integrated flux density spectrum}

\begin{figure}[!htbp]
    \centering
    \includegraphics[width=0.98\columnwidth]{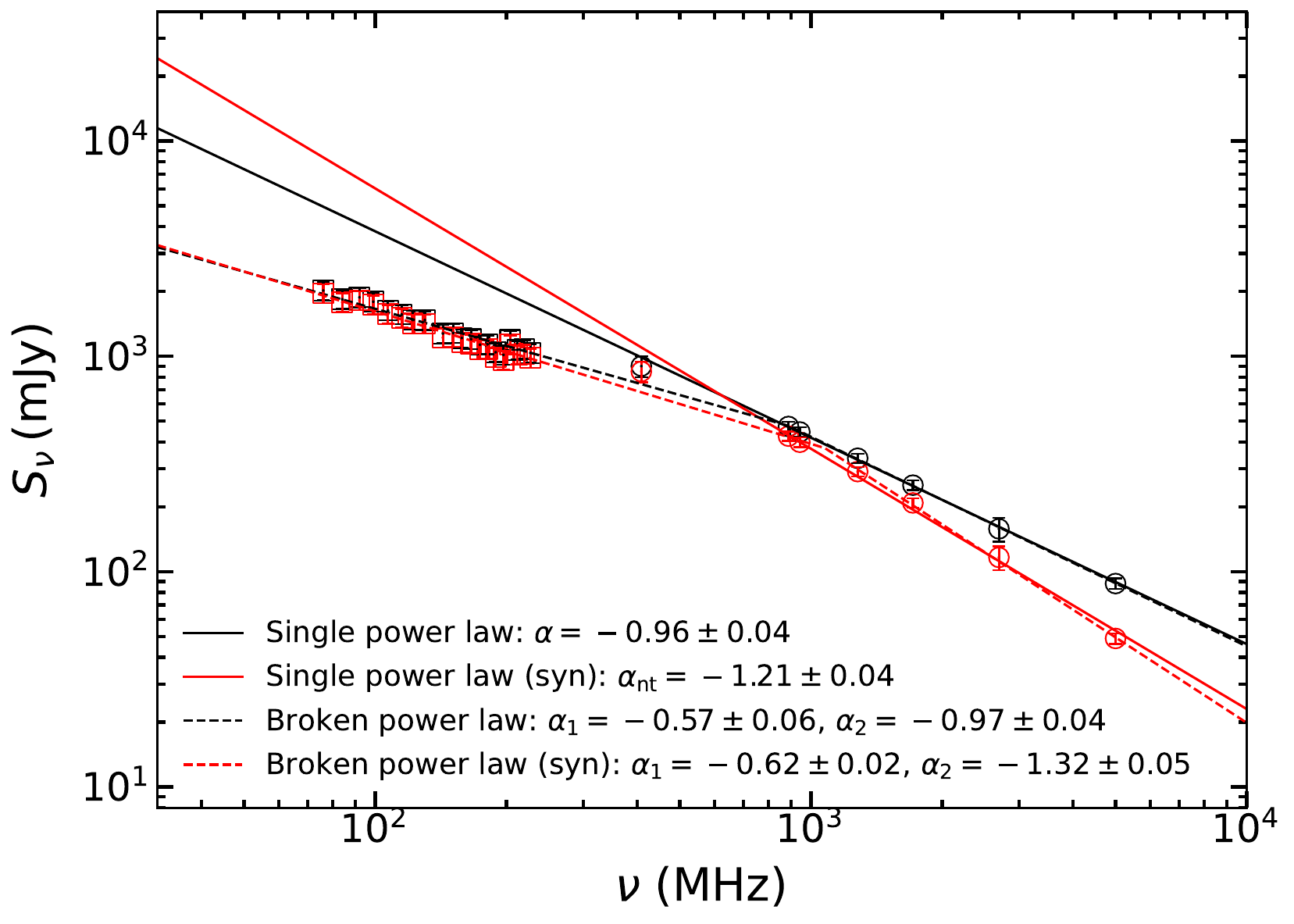}
    \caption{Integrated flux density spectrum of NGC~2442. The black and red symbols represent the integrated flux densities before and after thermal subtraction, respectively, with squares indicating the GLEAM measurements and circles indicating the measurements from the other surveys and instruments. The solid lines show the power-law fits for 408~MHz~$\leq\nu\leq$~5~GHz, while the dashed lines show the broken power-law fits to all data points. The black and red lines correspond to the fits to the total and synchrotron intensities, respectively.} 
\label{fig:spectral_index_fit}
\end{figure}

We fitted the data in Table~\ref{tab:int_flux} over the frequency range 408~MHz--5~GHz with a single power-law model, $S_\nu \propto \nu^{\alpha}$, using a least-squares fitting method, where $\alpha$ is the spectral index. The fits are shown in Fig.~\ref{fig:spectral_index_fit}, where the black and red solid lines represent the results for the total integrated flux density and the synchrotron emission, respectively, with corresponding spectral indices of $\alpha = -0.96\pm0.04$ and $\alpha_{\mathrm{nt}} = -1.21\pm0.04$. 

At frequencies $\nu < 408$~MHz, the spectrum flattens relative to the high-frequency single power-law fit, but shows no clear spectral inversion. We therefore fitted all data points with a broken power-law model to characterize the change in spectral slope. In logarithmic form, the model is written as~\citep{tisanic2019}
\begin{equation}
\log S_\nu =
\begin{cases}
\alpha_2 \log\left(\dfrac{\nu}{\nu_0}\right) + b, & \nu > \nu_{\rm b}, \\[12pt]
\alpha_1 \log\left(\dfrac{\nu}{\nu_0}\right) + b + (\alpha_2-\alpha_1)\log\left(\dfrac{\nu_{\rm b}}{\nu_0}\right), & \nu \leq \nu_{\rm b},
\end{cases}
\label{eq:bpl}
\end{equation}
where $\alpha_1$ and $\alpha_2$ are the spectral indices below and above the break frequency $\nu_{\rm b}$, respectively, $b$ is the normalization parameter, and $\nu_0$ is the reference frequency.

The dashed curves in Fig.~\ref{fig:spectral_index_fit} show the best-fitting broken power-law models for all data points. For the total integrated flux density, the fit yields a break frequency of $\nu_{\rm b} = 881 \pm 239$~MHz, with spectral indices of $\alpha_1 = -0.57 \pm 0.06$ below the break and $\alpha_2 = -0.97 \pm 0.04$ above it. For the synchrotron emission, the best-fitting model gives $\nu_{\rm b} = 1076 \pm 92$~MHz, with $\alpha_1 = -0.62 \pm 0.02$ and $\alpha_2 = -1.32 \pm 0.05$ below and above the break, respectively.

The broken power-law fit suggests that the integrated radio spectrum of NGC~2442 cannot be described by a single power law over the full frequency range. The flatter low-frequency spectrum may reflect the combined contribution of relatively younger and older CRE populations, while additional processes such as free--free absorption or ionization losses cannot be ruled out. In contrast, the steeper high-frequency spectrum indicates stronger radiative losses of higher-energy CREs through synchrotron and IC processes. The more pronounced steepening in the synchrotron spectrum after thermal subtraction further supports this interpretation, as the non-thermal component is expected to respond more directly to CRE aging. In this sense, the break frequency provides a characteristic scale at which radiative losses begin to modify the integrated spectrum significantly.

The derived high-frequency spectral indices for NGC~2442 ($\alpha \approx -1.0$ for the total emission and $\alpha_{\mathrm{nt}} \approx -1.2$ for the non-thermal component) are steeper than the typical values of $\alpha \sim -0.7$ to $-0.8$ commonly reported for nearby spiral galaxies~\citep{gioia1982,condon1992,paladino2009,tabatabaei2017}. This suggests that the global synchrotron emission of NGC~2442 is dominated by an evolved CRE population that has undergone substantial radiative losses. However, the integrated spectrum alone does not necessarily constrain the CRE transport properties, since it averages over regions with different CRE ages and ISM conditions. Stronger constraints on the transport physics are provided by the spatially resolved spectral-index distribution, as discussed below.

\subsubsection{Spectral index map}

\begin{figure*}
\sidecaption
\includegraphics[width=12cm]{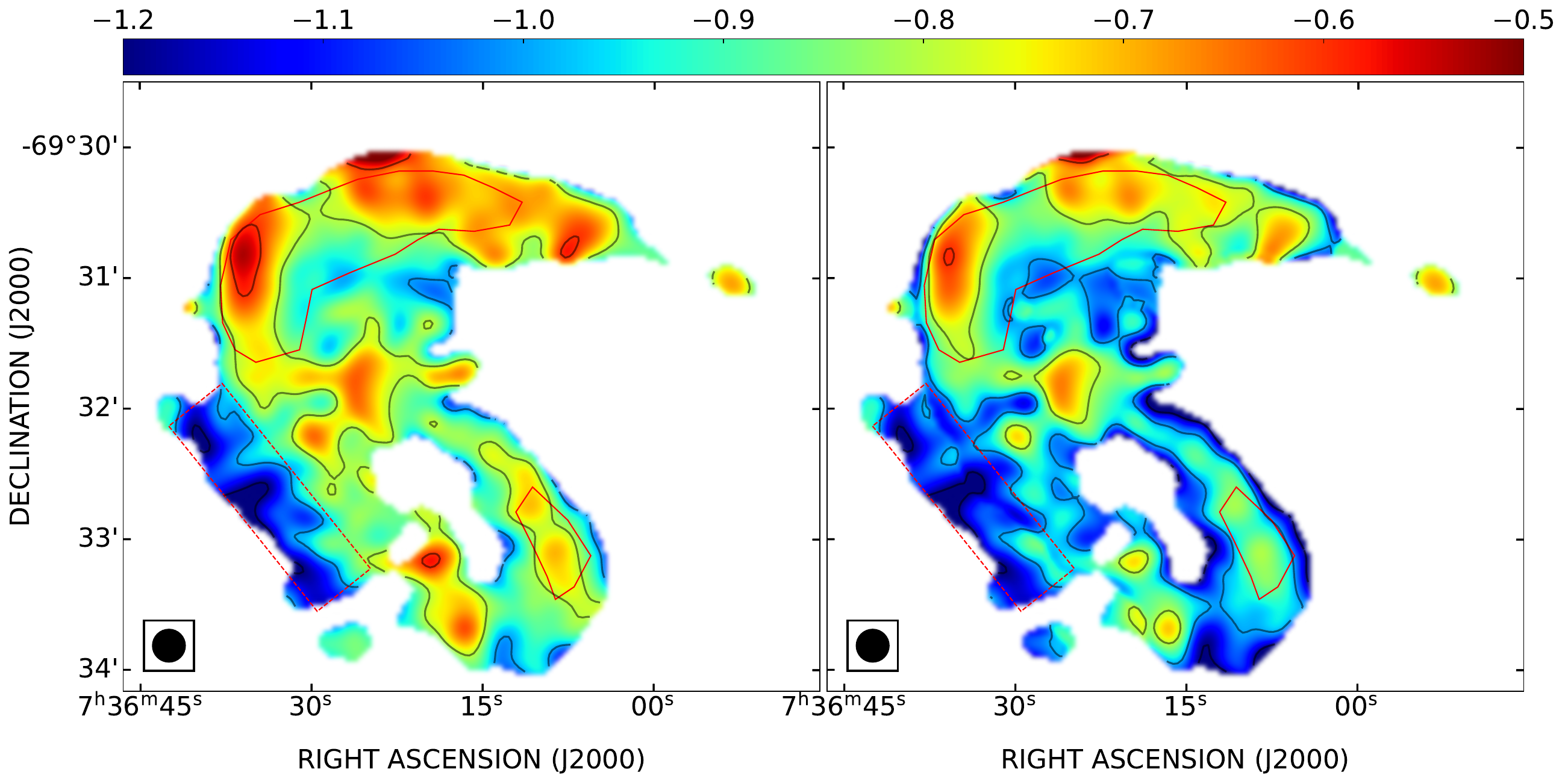}
\caption{Spectral index map for total intensity (left) and synchrotron emission (right). The black circles in the lower left corner indicate the beam size of $15\arcsec$. The black contour levels are $-$1.2, $-$1.0, $-$0.8, and $-$0.6. The red dashed rectangle outlines the island region, while the solid red contours outline the regions of high star formation activity.}
\label{fig:spex_distrib}
\end{figure*}

We derived a pixel-based spectral index map for total intensity~(Fig.~\ref{fig:spex_distrib}, left) using the ASKAP image at 943~MHz and the MeerKAT images at 1.28~GHz, 1.7~GHz and the ATCA at 5~GHz. We subtracted thermal emission pixel by pixel based on the estimate in Sect.~\ref{sec:therm_em} and derived the spectral index map for synchrotron emission (Fig.~\ref{fig:spex_distrib}, right panel). The spectral index is derived only when all the intensities at the four frequencies are larger than three times the rms noise level.

The synchrotron spectral index is systematically steeper than that of the total intensity, as expected from the contribution of thermal emission with a flatter spectrum to the total intensity. Even within the spiral arm, where star formation is active, NGC~2442 exhibits a steeper spectrum than most nearby star-forming galaxies. In the high SFR surface density ($\Sigma_{\rm SFR}$) region enclosed by the red solid contour in Fig.~\ref{fig:spex_distrib}, the mean spectral indices of the total intensity and synchrotron emission are $-0.78$ and $-0.82$, respectively. This may reflect the combined effects of the distorted arm geometry, enhanced energy losses, and local environment disturbance. The southeastern island region, marked by the red dashed rectangle in Fig.~\ref{fig:spex_distrib}, exhibits steep spectral indices, with average values of $-1.06$ for the total intensity and $-1.09$ for the synchrotron emission. This steepening likely reflects significant spectral aging of CREs due to strong energy losses during their propagation. Further discussion is provided in Sect.~\ref{subsec:origin_island}.

\subsection{Magnetic field strength}
\label{sec:equi_B}

\begin{figure}[!htbp]	
    \centering    \includegraphics[width=0.98\columnwidth]{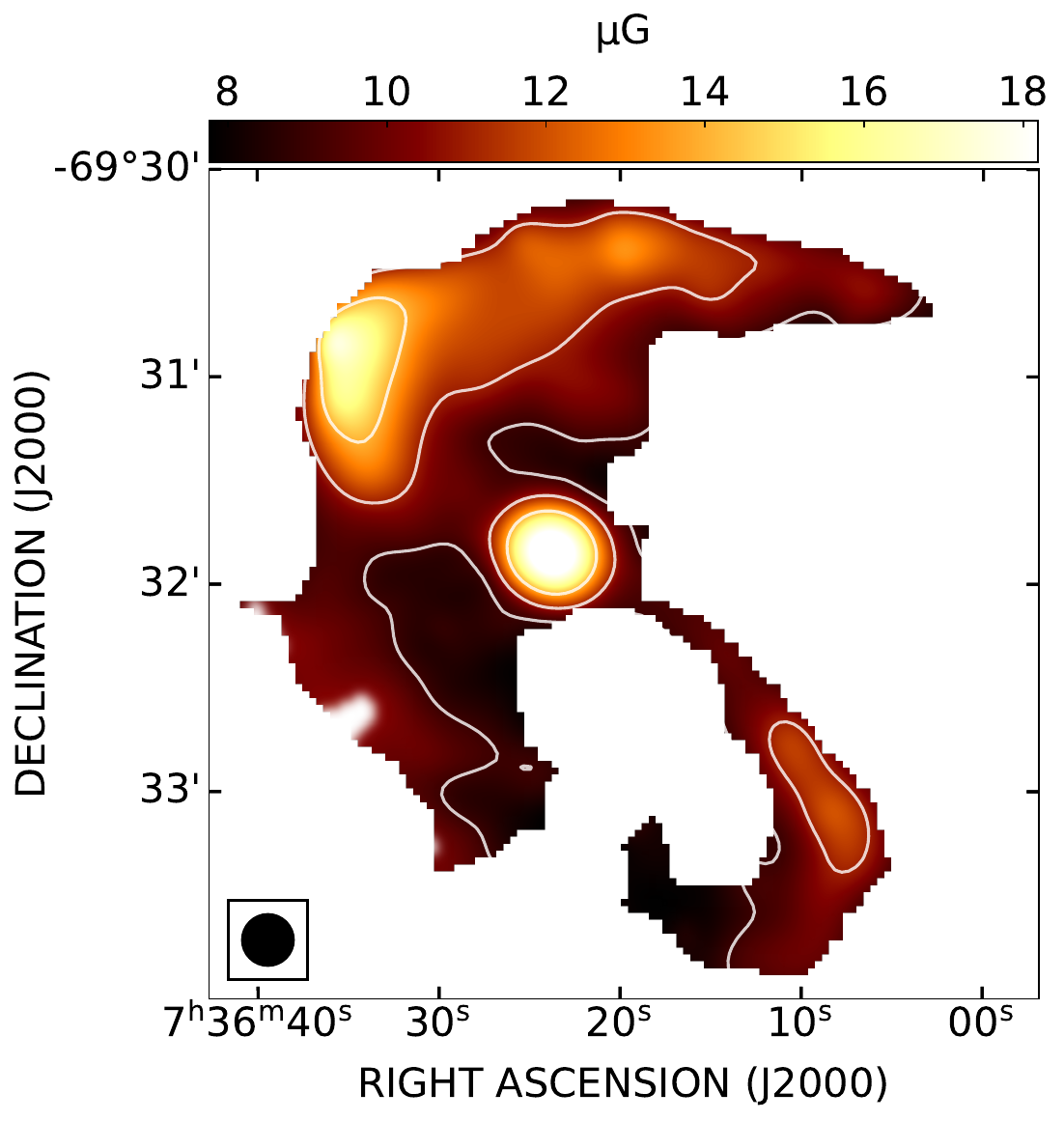}
    \caption{Total magnetic-field strength map derived under the assumption of energy equipartition. Contours are shown at 6, 8, 10, and 14~$\rm \mu G$. The black circle in the lower left corner represents the $15\arcsec$ beam.} 
\label{fig:B_distri}
\end{figure}

We estimated the magnetic field strength under the assumption of energy densities equipartition between CRs and the magnetic field \citep{beck2005}:
\begin{equation}
	B_{\mathrm{eq}} = \left( \frac{ 4\pi (2\alpha - 1) (K_0 + 1) I_\nu  E_p^{1+2\alpha} \left(\frac{\nu}{2c_1}\right)^{-\alpha} }{ (2\alpha + 1) c_2(\alpha) L \cdot c_4(i) } \right)^{\frac{1}{3 - \alpha}}, \label{eqn:B_eq}
\end{equation}
where $c_1$ is a constant, $c_2$ depends on the synchrotron emission spectral index $\alpha$, $E_p$ is the proton rest energy ($E_p=938.28\,\rm MeV=1.5\times 10^{-3}\,\rm erg$), $I_{\nu}$ is the synchrotron intensity in $\rm erg\,s^{-1}\,cm^{-2}\,Hz^{-1}\,Sr^{-1}$, $L$ is the effective path length through the source, $K_0$ is the ratio of the number densities of CR protons and electrons per particle energy interval over the energy range relevant for the observed synchrotron emission, and $c_4=(2/3)^{(1-\alpha)/2}$ assuming an isotropic magnetic field. 

To derive the magnetic field strength map, we used the ASKAP synchrotron emission intensity image, together with the synchrotron emission spectral index map (Fig.~\ref{fig:spex_distrib}, right panel). A spatially resolved value of $K_0$ would allow a more accurate estimate of $B_{\rm eq}$. However, $K_0$ cannot be constrained independently from the RC data alone, because this would require additional information about the CR proton population. We therefore adopt the standard value $K_0 = 100$, commonly used for normal star-forming disks, and regard possible spatial variations in $K_0$ as a systematic uncertainty. Since $B_{\rm eq}\propto(K_0+1)^{1/(3-\alpha)}$, a larger true $K_0$ would lead to a higher inferred field strength. Such an increase in $K_0$ is expected in regions where CREs have suffered stronger energy losses than CR protons, for example, in environments with strong synchrotron, IC, ionization, or bremsstrahlung losses, and in regions far from CRE injection sites such as outer disks and halos~\citep{seta2019}. This effect may be particularly relevant in the island region of NGC~2442, where the steep synchrotron spectrum and absence of local star-formation tracers suggest that CREs have propagated from elsewhere. The derived $B_{\rm eq}$ in this region may therefore be underestimated and should be regarded as a conservative estimate.

We adopted an effective path length of $L = 2h_{\rm syn}f$, where $h_{\rm syn}$ denotes the synchrotron scale height and $f$ is the volume filling factor of the synchrotron-emitting medium. For a nearly face-on galaxy, the line of sight is approximately parallel to the vertical direction of the disk. If the synchrotron emissivity decreases exponentially with height, $j_\nu(z)=j_{\nu,0}\exp(-|z|/h_{\rm syn})$, the observed synchrotron surface brightness is obtained by integrating the emissivity along the line of sight:
\begin{equation}
   I_\nu=\int_{-\infty}^{+\infty}f\,j_\nu(z)\,dz=2f\,j_{\nu,0}\,h_{\rm syn}. 
\end{equation}
Thus, an exponential emitting layer has the same integrated surface brightness as a uniform layer with emissivity $j_{\nu,0}$ and effective path length $L=2h_{\rm syn}f$. In this sense, $L$ represents an effective
emissivity-weighted path length rather than the full physical size of
the radio halo. For most nearby edge-on spiral galaxies, a typical synchrotron scale height of $\sim 1.5$~kpc is found~\citep{krause2018,heesen2025}, corresponding to an effective path length of $L \approx 3$~kpc for the adopted filling factor $f=1$, assuming that the diffuse synchrotron-emitting medium approximately fills the relevant volume on the spatial scales considered here. Note that $B_{\rm eq}$ is only weakly dependent on $f$. To avoid overestimating the magnetic field strength, we used only regions with spectral indices in the range $-1.2 < \alpha < -0.6$, following the discussion in~\citet{heesen2022}. The resulting distribution of the magnetic-field strength is shown in Fig.~\ref{fig:B_distri}. The mean magnetic-field strength is $10.8~\mu{\rm G}$, which is comparable to the typical total magnetic-field strengths of nearby spiral galaxies, usually of order $10~\mu{\rm G}$~\citep{beck2000,beck2015,krause2018,heesen2022}. Previous studies have shown that total, mostly turbulent magnetic fields are enhanced in spiral arms and bars, whereas ordered fields are often stronger in interarm regions or along the inner edges of spiral arms~\citep{beck2015}. Therefore, NGC~2442 is not exceptional in terms of its galaxy-averaged magnetic-field strength. Its peculiarity instead appears to be associated with the disturbed gas and star-formation morphology, together with the ordered magnetic-field geometry that may guide CREs from the star-forming disk toward the southeastern synchrotron island. The possible origin of this island is discussed further in Sect.~\ref{subsec:origin_island}.

\subsection{RC-SFR relation in NGC~2442}
\label{subsec:CRE_trans}
\subsubsection{SFR surface density maps}

Following \citet{leroy2008}, we estimate the star formation rate using the hybrid FUV and 24~$\mu$m method:
\begin{equation}
(\Sigma_{\rm SFR})_{\rm hyb}
= \left(8.1\times10^{-2}\,I_{\rm FUV}+3.2\times10^{-3}\,I_{24}\right)\,\mathrm{cos}\,i,
\label{eq:hyb_SFR}
\end{equation}
where $(\Sigma_{\rm SFR})_{\rm hyb}$ has units of $\mathrm{M_\odot\,yr^{-1}\,kpc^{-2}}$, the FUV and $24\,\mu$m intensities are both expressed in $\rm MJy\,Sr^{-1}$, and $i$ is the inclination angle of the galaxy. The FUV traces young star formation through the UV radiation from massive stars. Dust efficiently absorbs this UV radiation and re-emits it in the infrared, so the 24~$\mu$m emission traces embedded star formation. Following \citet{condon1992} and \citet{heesen2014}, the relation between the RC emission and the SFR is given by

\begin{equation}
\begin{aligned}
\frac{\left(\Sigma_{\mathrm{SFR}}\right)_{\mathrm{RC}}}
{\mathrm{M_\odot\,yr^{-1}\,kpc^{-2}}}
&= 3.31 \times 10^{3}
\left( \frac{\nu}{1.4\,\mathrm{GHz}} \right)^{0.8} \\
&\quad \times
\left( \frac{\mathrm{FWHM}}{\mathrm{arcsec}} \right)^{-2}
\frac{I_\nu}{\mathrm{Jy\,beam^{-1}}}\cos i .
\end{aligned}
\label{eq:RC_SFR}
\end{equation}

\begin{figure*}[!htbp]	
    \centering    \includegraphics[width=\linewidth]{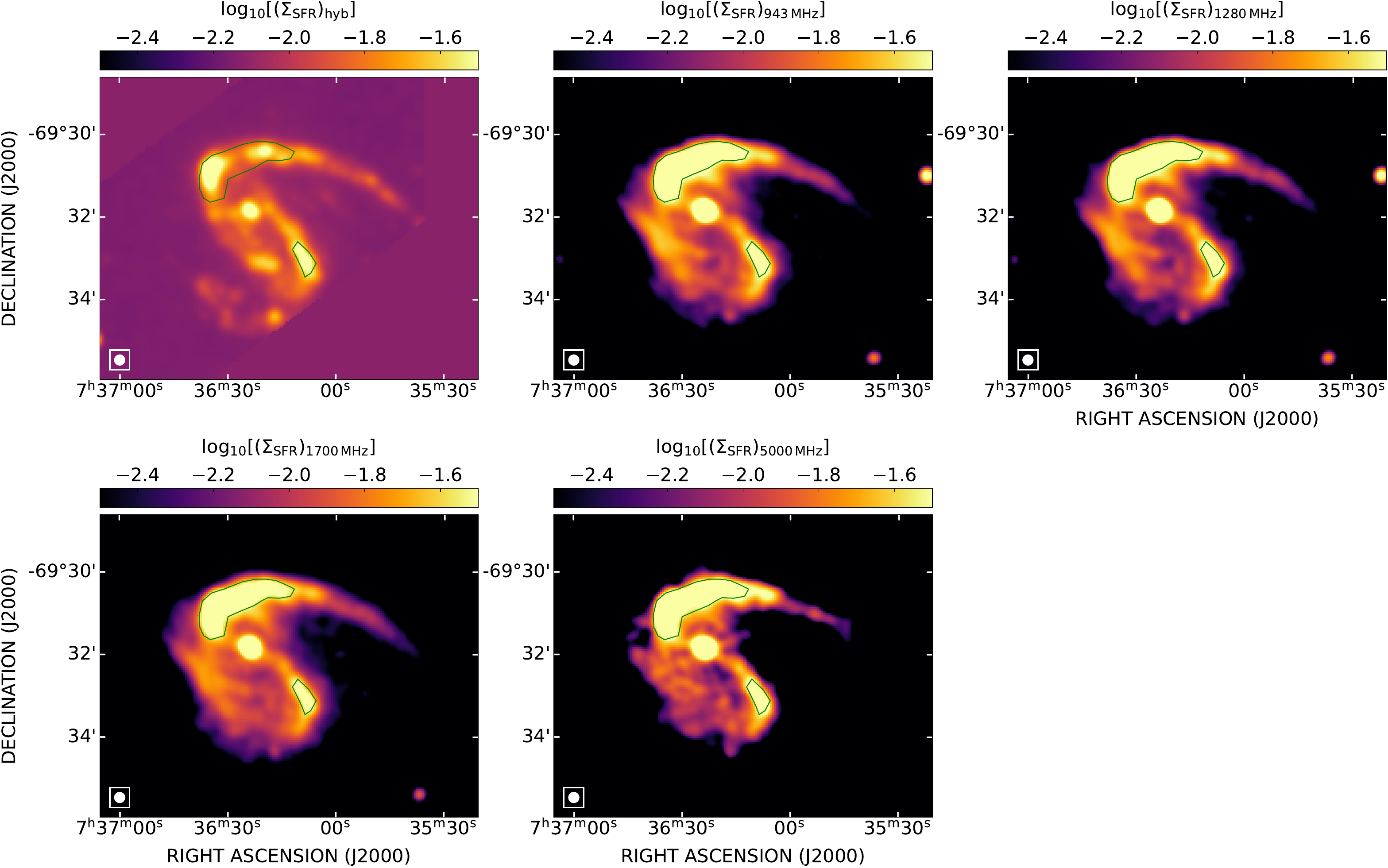}
    \caption{SFR surface density maps of NGC~2442 derived using the hybrid method and RC measurements at 943, 1280, 1700 and 5000~MHz. The region outlined by the green contour is used for the smoothing experiment.} 
\label{fig:SFR_surface_density}
\end{figure*}

All maps are convolved with a Gaussian kernel to a common angular resolution of $15\arcsec$ and regridded to a common coordinate system. Based on Eqs.~(\ref{eq:hyb_SFR}) and (\ref{eq:RC_SFR}), we computed the SFR surface density of NGC~2442, as shown in Fig.~\ref{fig:SFR_surface_density}. The spiral-arm region outlined by the green contour corresponds to an area of enhanced $\Sigma_{\rm SFR}$, with values ranging from $0.02$ to 0.05~$\rm M_{\odot}\,yr^{-1}\,kpc^{-2}$.

The radio--infrared and radio--SFR relations are known to be tight on global scales~\citep{condon1992,yun2001,li2016}, but their behavior on sub-galactic scales is affected by CRE propagation away from star-forming regions. The infrared and hybrid SFR tracers respond more directly to recent massive star formation, whereas non-thermal radio emission is spatially broadened by CRE diffusion or advection before the electrons lose their energy. This effect has been used to estimate characteristic CRE propagation lengths by smoothing infrared or SFR maps until they best match the radio continuum distribution~\citep[e.g.][]{bicay1990,murphy2006,murphy2008,
heesen2014}.

\subsubsection{Smoothing experiment}
\label{subsec:smooth_exp}
\begin{figure}[!htbp]	
    \centering    \includegraphics[width=0.98\columnwidth]{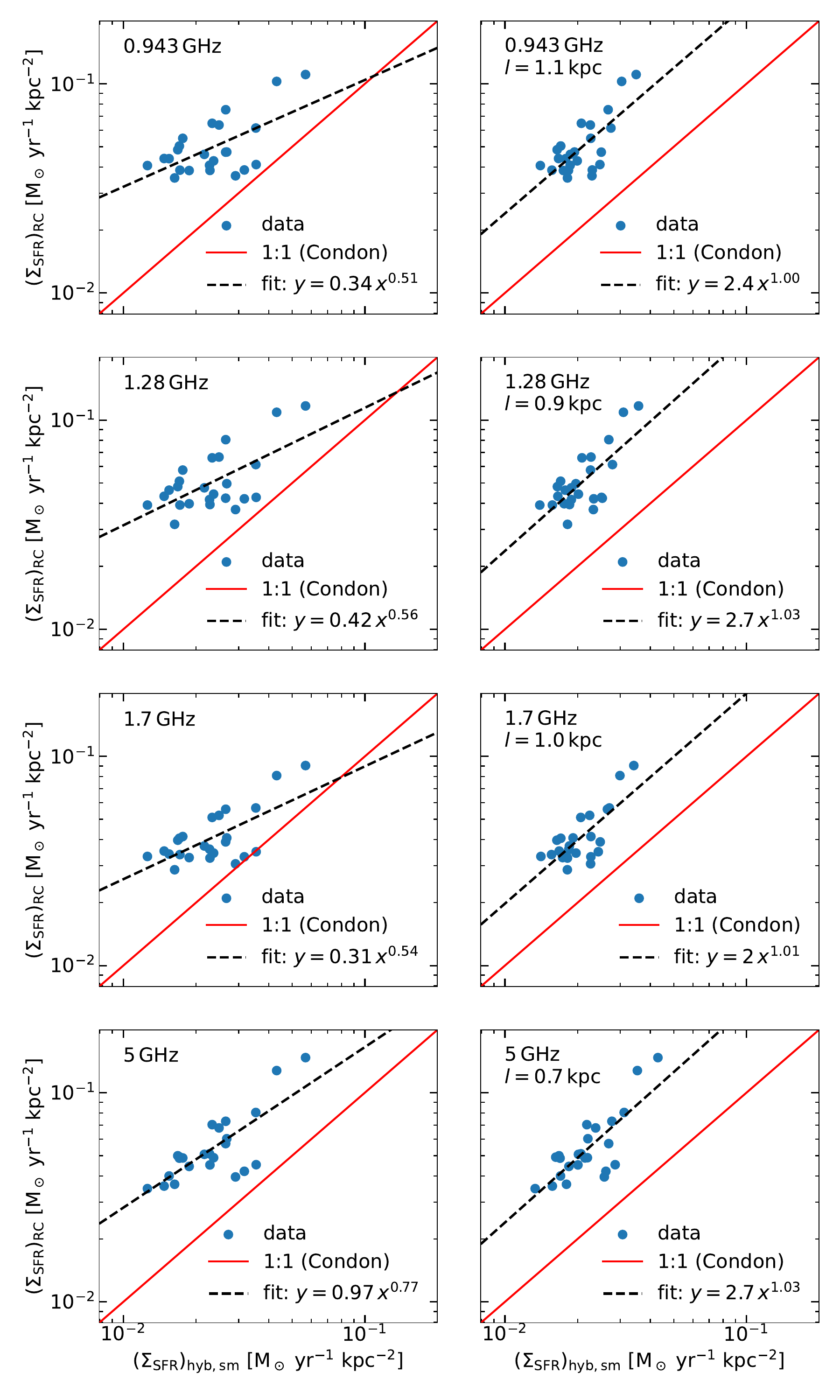}
    \caption{Linearized $\rm (\Sigma_{SFR})_{RC}$–$\rm (\Sigma_{SFR})_{hyb}$ relation after convolving the $\rm (\Sigma_{SFR})_{hyb}$ map with a Gaussian kernel to simulate the diffusion of CREs. The left and right panels show the relations before and after convolution, respectively. The red solid line shows the Condon 1:1 relation, and the black dashed line shows the linear fit. The observing frequency and the convolution scale $l$ are also indicated.} 
\label{fig:smooth_exp_sfr}
\end{figure}

 We expect advection to be important primarily in the halo, whereas the emission analyzed here arises mainly from the thin disk. Diffusion is therefore expected to be the dominant CR transport process in the disk. By convolving the $\rm (\Sigma_{SFR})_{hyb}$ map with an appropriate Gaussian kernel, we can linearize the RC--SFR relation and estimate the CRE transport scale. Fig.~\ref{fig:smooth_exp_sfr} shows the relation between $\rm (\Sigma_{SFR})_{RC}$ and $\rm (\Sigma_{SFR})_{hyb}$ before and after smoothing, in the left and right panels, respectively. The diffusion kernel length, $l$, is defined as half the full width at half maximum (FWHM) of the Gaussian convolution kernel applied to the hybrid $\Sigma_{\rm SFR}$ map, i.e. $l =\mathrm{FWHM}/2$. The CRE diffusion length is then given by $l_{\rm CRE}^2 = l^2 - l_{\rm beam}^2$, where $l_{\rm beam} = 7\farcs5$. The resulting diffusion lengths are listed in Table~\ref{tab:smooth_exper}.

We can calculate the diffusion coefficient using the simplified equation $D=l_{\rm CRE}^2/\tau$, where $\tau$ is the CRE lifetime due to synchrotron and IC radiation losses~\citep{heesen2016}:
\begin{equation}
\tau = 34.2 \left (\frac{\nu}{\rm 1~GHz}\right )^{-0.5}
\left (\frac{B_\perp}{\rm 10~\mu G}\right )^{-1.5} \left
  (1+\frac{U_{\rm rad}}{U_{\rm B}}\right )^{-1}~{\rm Myr},
\label{eq:t_rad}
\end{equation}
where $U_{\rm B} = B^2/(8\pi)$ is the magnetic energy density and $U_{\rm rad}=U_{\rm CMB}+U_{\rm TIR}+U_{\rm \star}$ is the total radiation energy density. The cosmic microwave background (CMB) radiation energy density is $U_{\rm CMB}=4.2\times10^{-13}\,\rm erg\,cm^{-3}$ at redshift zero~\citep{heesen2016}. The starlight radiation energy density is related to the total infrared radiation energy density by $U_{\rm \star}=1.73\,U_{\rm TIR}$. The latter is given by $U_{\rm TIR}=L_{\rm TIR}/(2\pi r_{\star}^2c)$, where $r_{\star}=7$~kpc is the radius of the actively star-forming disk, estimated from the $(\Sigma_{\rm SFR})_{\rm hyb}$ map, and $L_{\rm TIR}$ is the total infrared luminosity listed in Table~\ref{tab:parameters_table}. We derive a total infrared radiation energy density of $8.7\times10^{-13}\,\rm erg\,cm^{-3}$, resulting in total radiation energy density of $2.8\times10^{-12}\,\rm erg\,cm^{-3}$. The spatially averaged magnetic field strength within the green region shown in Fig.~\ref{fig:SFR_surface_density} is $12.4\,\rm \mu G$. Finally, we estimated the CRE lifetimes and diffusion coefficients at different frequencies, as listed in Table~\ref{tab:smooth_exper}.

The smoothing analysis yields CRE propagation lengths of $\sim0.65-0.89$~kpc at $943-1700$~MHz, while a smaller value of $\sim0.44$~kpc is found at 5~GHz. The results at $943-1700$~MHz are broadly consistent with each other, suggesting that CRE transport on kiloparsec scales can be approximately described by diffusion with an effective coefficient of order $10^{28}\,\mathrm{cm^2\,s^{-1}}$. The lower value derived at 5~GHz is likely related to the shorter lifetime of higher-energy CREs and the fact that high-frequency radio emission is more strongly weighted toward younger, more localized CRE populations. In addition, thermal contamination and local variations in the magnetic field and star-forming environment may also affect the smoothing-derived propagation scale at 5~GHz. The diffusion coefficients derived here should be regarded as effective, spatially averaged isotropic diffusion coefficients. Since CRE transport is strongly influenced by the anisotropic magnetic-field structure, these values mainly provide a global characterization of the overall transport efficiency. A more detailed discussion is given in Sect.~\ref{subsec:origin_island}.

\begin{table}[!htbp]
\caption{Properties of CRE diffusion.}
\centering
\label{tab:smooth_exper}
\begin{tabular}{ccccc}     
\hline\hline
Frequency & $l$  & $l_{\rm CRE}$ & $\tau$ & $D$ \\ 
(MHz)     & (pc) & (pc)          & (Myr)  & ($\rm 10^{28}\,cm^2\,s^{-1}$)\\
\hline
943       & 1050 & 886           & 23.2   &1.02\\
1280      & 863  & 654           & 20.0   &0.65\\      
1700      & 1012 & 841           & 17.3   &1.23\\
5000      & 713  & 437           & 10.1   &0.57\\ 
\hline
\end{tabular}
\end{table}

\section{Discussion}
\label{sec:discussions}

\subsection{2D CRE propagation simulation in NGC~2442}
\label{subsec:3D_CRE_propa}

\begin{figure}[!htbp]	
    \centering    \includegraphics[width=0.98\columnwidth]{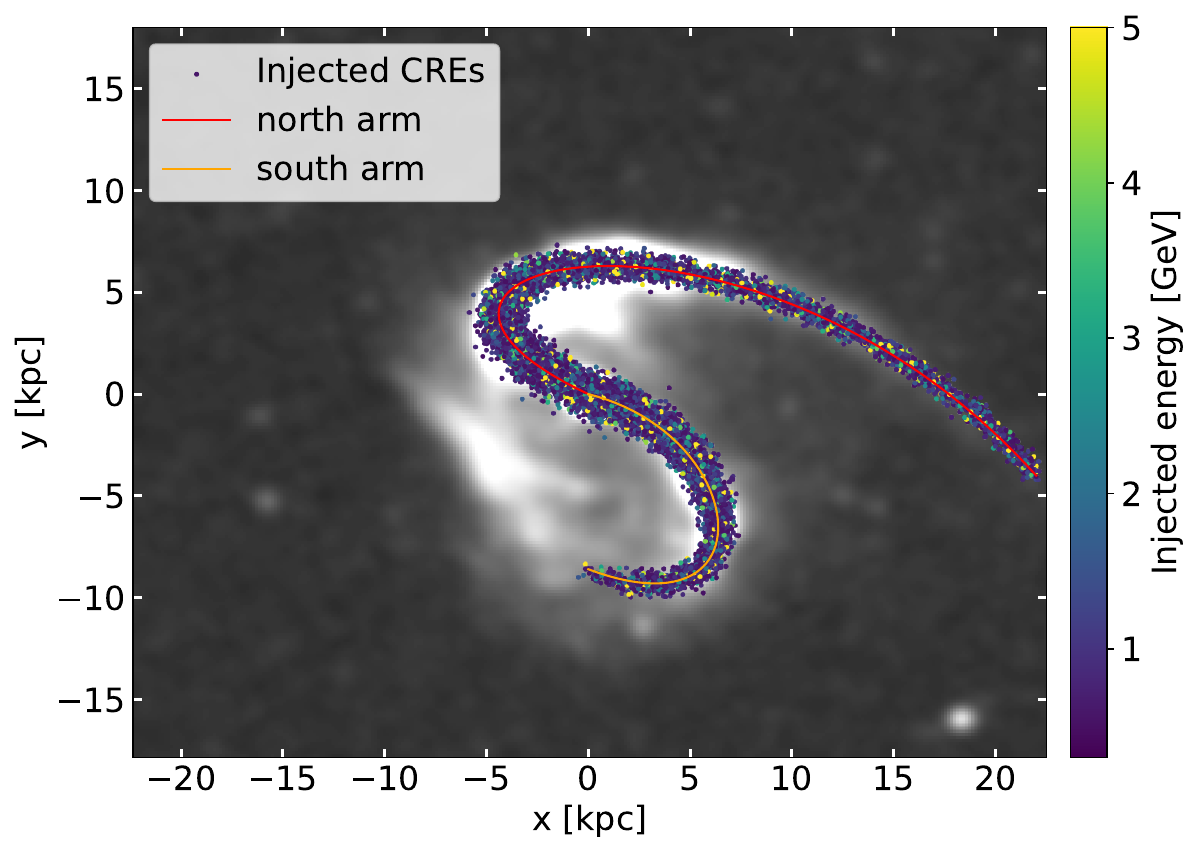}
    \caption{The two spiral-arm structures of NGC~2442 and the assumed CRE injection sites along the arms. The grayscale background shows the total intensity map at 943 MHz.} 
\label{fig:CRE_model}
\end{figure}

To model the 2D transport of CREs in the disk plane of NGC~2442, we assume that diffusion is the dominant transport mechanism and that the particles lose energy during propagation. The evolution of the CRE distribution is described by~\citep{berezinskii1990}

\begin{equation}
\label{adv_diff}
\frac{\partial N}{\partial t}
= D(E)\nabla^{2}N
+ \frac{\partial}{\partial E}\left[b(E)N\right]
+ Q(\mathbf{r},E)
- \frac{N}{\tau} \, .
\end{equation}
The first term on the right-hand side describes spatial diffusion, where $D(E)$ is the diffusion coefficient and $\nabla^{2}N$ is the Laplacian of $N$. The second term represents energy losses, with $b(E) = -\,dE/dt$ denoting the energy loss rate. The third term is the source term, describing the injection rate of particles as a function of position $r$ and energy $E$. The final term accounts for escape or other losses, where $\tau$ is the characteristic timescale for CRE removal. We study the 2D transport of CREs in NGC~2442 using the publicly available Monte Carlo code \texttt{CRPropa} version~3.2.1\footnote{\url{https://crpropa.desy.de}}~\citep{merten2017,alves2022,dorner2023,aravinthan2025}.

We modeled the two spiral arms of NGC~2442 using B\'ezier curves and assumed that CREs are injected along them. The injected CRE number density was assumed to decline exponentially in the direction perpendicular to the arms, from the inner galaxy toward the outer spiral-arm ends. We injected a total of \(3\times10^4\) CREs with energies between 0.5 and 20~GeV, following a power-law spectrum, \(dN/dE \propto E^{-\gamma}\), with an injection index of \(\gamma_{\rm inj}=2.1\), consistent with the range commonly adopted for particle acceleration in supernova remnants~\citep{blandford1987,dubner2015}. The initial CRE distribution is shown in Fig.~\ref{fig:CRE_model}. We modeled the 2D magnetic-field structure of NGC~2442 by assuming that the $x$--$y$ plane represents the face-on view of the galaxy. In this plane, the field lines follow the spiral-arm pattern, similar to that indicated by the ATCA at 5~GHz polarization observations~\citep{harnett2004}. The resulting magnetic-field model for NGC~2442 is shown in Fig.~\ref{fig:ngc2442_2D_Bfield}, while the detailed modeling procedure is presented in Appendix~\ref{sec:B_model}.

\begin{figure}[!htbp]	
    \centering    \includegraphics[width=\linewidth]{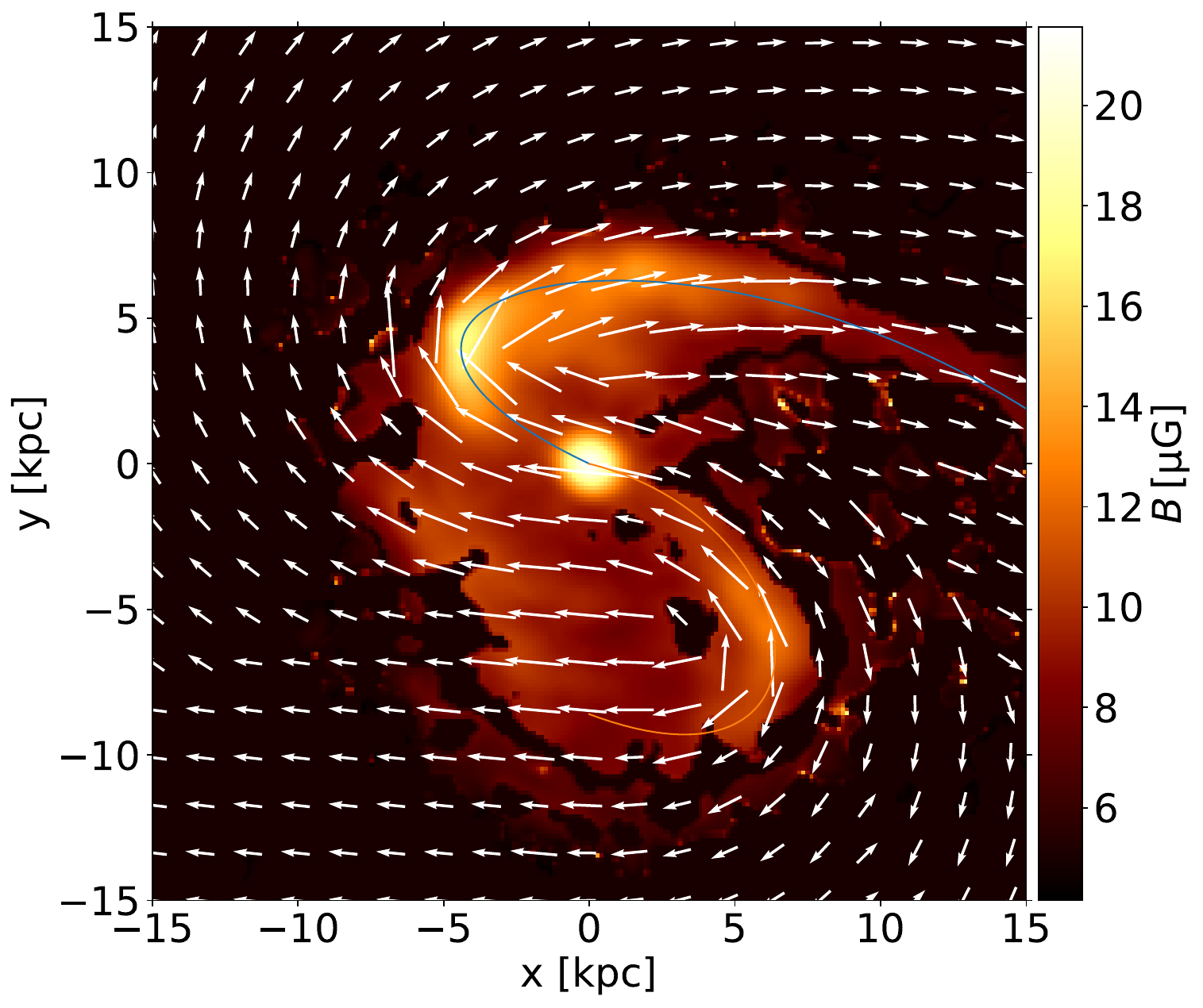}
    \caption{The magnetic-field orientations of NGC~2442 for $\theta = 10\degr$ are overlaid on the equipartition magnetic-field strength map shown in Fig.~\ref{fig:B_distri}.} 
\label{fig:ngc2442_2D_Bfield}
\end{figure}

\begin{figure*}[!htbp]	
    \centering    \includegraphics[width=\linewidth]{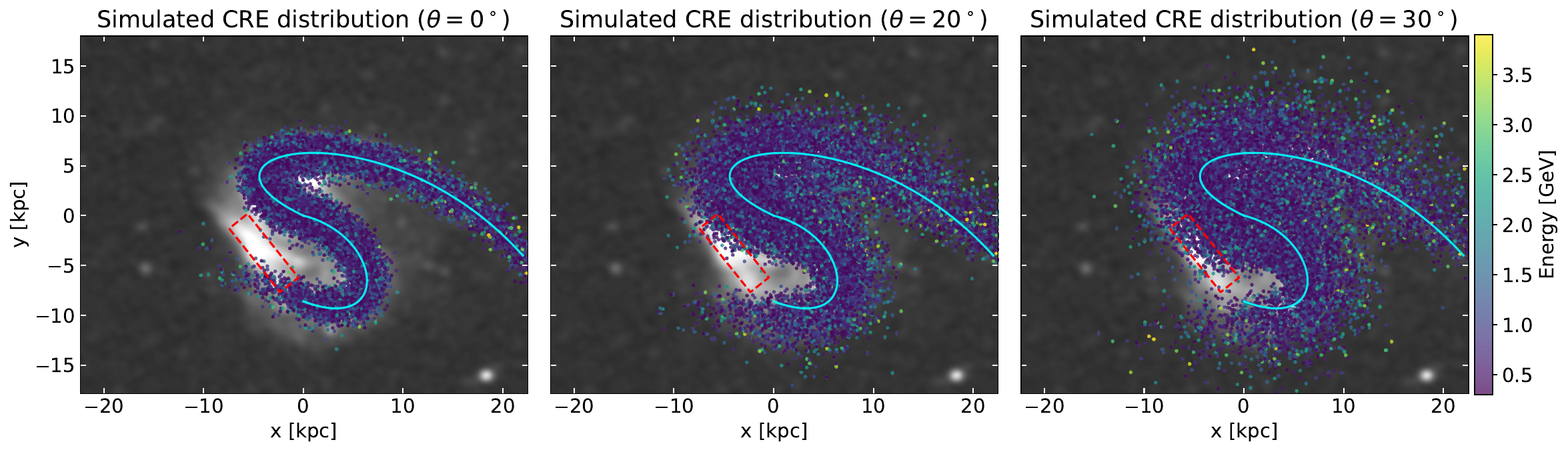}
    \caption{Simulated distributions of diffusing CREs in the $x$--$y$ plane after 23.2~Myr. From left to right, the panels show the propagation results for different magnetic-field angles, $\theta$. The background shows the 943~MHz total-intensity image. The cyan curves indicate the two spiral arms of NGC~2442, and the red dashed rectangle marks the island region.} 
\label{fig:CREs_sim}
\end{figure*}

To account for anisotropic diffusion, we adopted a ratio of 40 between the diffusion coefficients parallel and perpendicular to the magnetic-field orientation, and set $D_{\parallel} = 3.0 \times 10^{29}\ {\rm cm^2\,s^{-1}}$. The propagation of these CREs was implemented in CRPropa over a timescale of 23.2~Myr. The resulting CRE distributions for different magnetic-field orientations $\theta$ are shown in Fig.~\ref{fig:CREs_sim}. We find that CREs can reach the island region more easily as $\theta$ increases. This behavior is expected, because a larger $\theta$ increases the in-disk magnetic-field component perpendicular to the spiral arms, thereby favoring outward radial diffusion of CREs. As shown in the right panel of Fig.~\ref{fig:CREs_sim}, the island region contains the largest number of CREs.

Therefore, the diffusion of CREs depends strongly on the magnetic-field orientation, with transport being more efficient along the field lines. Compared to the isotropic diffusion case presented in Sect.~\ref{subsec:smooth_exp}, anisotropic diffusion allows CREs to reach the island region more easily. In the right panel of Fig.~\ref{fig:CREs_sim}, the island region has an electron spectral index of $\gamma=2.88$, corresponding to a synchrotron spectral index of $\alpha=-(\gamma-1)/2=-0.94$, which is slightly flatter than the value of $\approx-1.1$ derived in Sect.~\ref{subsec:spex}. This discrepancy is likely due to the limitations of our magnetic-field model. Observations at 5~GHz reveal that the island region exhibits a highly ordered magnetic field aligned along the northeast-southwest direction of the galaxy, with a very high linear fractional polarization of $\approx52\%$~\citep{harnett2004}. This suggests that once the CREs propagate into this region, they will be strongly confined by the magnetic field, leading to more rapid synchrotron cooling and consequently resulting in a steeper spectral index. In addition, in strongly curved arm segments, the field orientation varies rapidly with position, causing CREs injected at nearby locations to follow different directions and produce a fan-like spread. Thus, the enhanced outward extension in the island region may partly reflect a geometric effect of the curved arm-aligned magnetic field, rather than a true local increase in the diffusion coefficient.

However, because of the limitations of current polarimetric observations, especially the strong Faraday depolarization affecting low-frequency L-band data, we are unable to constrain the small-scale magnetic-field distribution, which would also affect CR transport. Moreover, a more realistic 3D magnetic-field model should provide tighter constraints on CRE propagation. In particular, including a magnetic-field component perpendicular to the galactic disk, that is, along the $z$ direction, may be important for describing the transport more accurately. Future high-sensitivity observations at higher frequencies, for example, with MeerKAT or the SKA in S band, will provide much stronger constraints on this model.

\subsection{Origin of the island}
\label{subsec:origin_island}
\subsubsection{Anisotropic diffusion of CREs}
\begin{figure}[!htbp]	
    \centering
    \includegraphics[width=0.98\columnwidth]{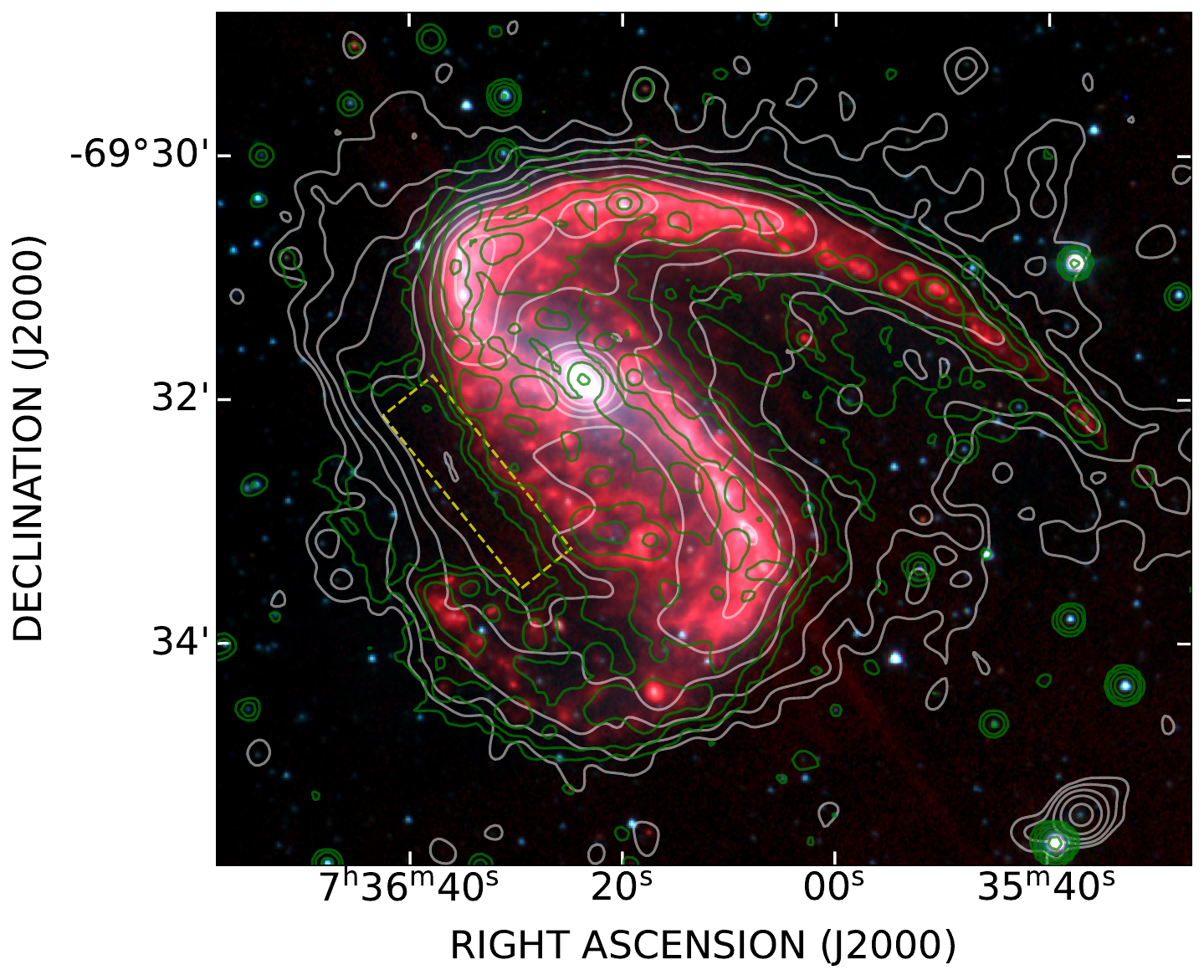}
    \caption{Three-color IRAC image of NGC~2442, with the 3.6, 4.5, and 8~$\mu$m bands shown in blue, green, and red, respectively. The white contours show the ASKAP 943~MHz RC intensity, and the green contours show the NUV intensity. Contours are drawn at levels of $3\sigma \times 2^n$ ($n=0,1,2,\ldots$), where the rms noise levels are $\sigma=25~\mu{\rm Jy~beam^{-1}}$ for the radio image and $\sigma=9\times10^{-4}~{\rm MJy~sr^{-1}}$ for the NUV image. The yellow dashed box outlines the island region.} 
\label{fig:NGC2442_RGB_map}
\end{figure}

Fig.~\ref{fig:NGC2442_RGB_map} shows the ASKAP 943~MHz total-intensity contours and the NUV contours overlaid on a three-color IRAC image,   with the 3.6, 4.5, and 8~$\mu$m bands shown in blue, green, and red, respectively. The island region, marked by the yellow dashed rectangle, shows no clear NUV counterpart and no obvious mid-infrared enhancement associated with ongoing star formation. This supports the interpretation that the CREs responsible for the synchrotron emission are unlikely to be injected in situ.

A purely isotropic transport picture is disfavoured by the characteristic propagation scale derived from the RC--SFR smoothing analysis. We find effective CRE propagation lengths of only $l_{\rm CRE}\sim0.65-0.89$~kpc at 943--1700 MHz and $l_{\rm CRE}\sim0.44$ kpc at 5 GHz, both substantially smaller than the observed arm--island separation of $\sim5$ kpc. Using the CRE lifetimes derived in this work, transport across 5 kpc corresponds to characteristic propagation speeds of order $V\sim 210\ {\rm km\ s^{-1}}$ at 943~MHz and $V\sim480\ {\rm km\ s^{-1}}$ at 5~GHz. However, these values are difficult to interpret as large-scale radial advection speeds, because such advection would require a sustained and coherent in-plane bulk flow over several kiloparsecs, for which there is no direct observational evidence in the island region. Moreover, unlike the vertical outflows commonly inferred in edge-on galaxies, the transport considered here occurs predominantly within the disk plane, where ordered magnetic fields provide a more natural channel for directional CRE propagation. We therefore argue that the island cannot be explained as a simple extension of the disk-averaged isotropic diffusion inferred from the smoothing analysis, and instead requires a more efficient and directional transport mechanism.

The most plausible interpretation is therefore that CREs are injected in the spiral-arm regions and transported into the island by anisotropic diffusion along an ordered magnetic field. This scenario is supported by the high fractional polarization ($\approx 52\%$), the highly ordered magnetic field, and the coherent RM structure reported by~\citet{harnett2004}, as well as by our CRPropa models. For anisotropic diffusion, the characteristic propagation lengths parallel and perpendicular to the magnetic field are $l_{\parallel}\simeq\sqrt{2D_{\parallel}\tau}$ and $l_{\perp}\simeq\sqrt{2D_{\perp}\tau}$, respectively. Using $D_{\parallel}=3\times10^{29}\ {\rm cm^2\,s^{-1}}$ and $D_{\parallel}/D_{\perp}=40$, we obtain $l_{\parallel}\approx 6.8$ kpc and $l_{\perp}\approx 1.1$~kpc at 943 MHz, and $l_{\parallel}\approx 4.5$ kpc and $l_{\perp}\approx 0.7$ kpc at 5 GHz. Thus, the observed arm--island separation of $\sim 5$ kpc can be reached naturally along ordered magnetic field lines, but is difficult to explain through cross-field diffusion. Cross-field processes such as field-line random walk, mirroring, or turbulent scattering may still broaden the CRE distribution locally~\citep[e.g.][]{jokipii1966,shalchi2007,shukurov2017,lazarian2023}, but they are unlikely to provide the dominant transport channel over the full arm--island separation.

\subsubsection{Interaction of NGC~2442 with its group environment}

We further suggest that the island is not merely the endpoint of CRE transport, but rather an environmentally shaped magnetic structure. Our CRPropa simulations demonstrate that anisotropic diffusion can naturally transport CREs from the spiral arm to the island over the observed separation, explaining the presence of synchrotron emission several kiloparsecs from the nearest star-forming region. However, CRE transport alone cannot readily explain the exceptionally high fractional polarization observed in the island, implying that the local magnetic-field geometry has also been modified.

NGC~2442 resides in a disturbed group environment, where both tidal interaction and ram pressure have previously been invoked to explain its asymmetric morphology. The associated {\sc H\,i} cloud HIPASS~J0731--69, containing $\sim10^9\,\rm M_\odot$ of atomic gas~\citep{ryder2001}, provides strong evidence for recent gas removal or displacement and therefore indicates that environmental processes have influenced the evolution of the system.

To assess whether ram pressure is capable of producing the highly polarized island, we compare the ram pressure with the local magnetic pressure. The magnetic pressure associated with the mean field strength in the island region is
$P_B=B_{\rm island}^2/8\pi \simeq4.2\times10^{-12}\,{\rm dyn\,cm^{-2}}$
for $B_{\rm island}=10.3\,\mu{\rm G}$. For a typical intragroup medium (IGM) density of
$n_{\rm IGM}\sim10^{-4}$--$10^{-3}~{\rm cm^{-3}}$~\citep{mulchaey2000,sun2012}, and using the
line-of-sight velocities of the neighboring galaxies listed by
\citet{ryder2001}, we estimate a group mean velocity of
$\bar{V}\simeq1400~{\rm km~s^{-1}}$ and a line-of-sight velocity
dispersion of $\sigma_{\rm los}\simeq70~{\rm km~s^{-1}}$. Assuming isotropic motions, this corresponds to a characteristic three-dimensional velocity of $V=\sqrt{3}\,\sigma_{\rm los}=120~{\rm km~s^{-1}}$. The corresponding ram pressure is estimated to be
\begin{equation}
P_{\rm ram}=\rho_{\rm IGM}\,V^2
\simeq 3\times10^{-14}-3\times10^{-13}
~{\rm dyn~cm^{-2}},
\end{equation}
where $\rho_{\rm IGM}=\mu_e\,n_{\rm IGM}\,m_p$, $\mu_e\simeq1.2$, and
$m_p$ is the proton mass. The estimated ram pressure is therefore approximately one to two orders of magnitude lower than the magnetic pressure in the island region, suggesting that ram pressure exerted by the diffuse IGM is unlikely to be the dominant mechanism responsible for producing the island or compressing its magnetic field.

Instead, this suggests that tidal interaction may play an important role in shaping the island. The magnetic-field strength in the island ($10.3\,\mu{\rm G}$) is very similar to the galaxy-wide average ($10.8\,\mu{\rm G}$), indicating that the total magnetic-field strength has not been significantly enhanced. This suggests that the observed radio enhancement is not driven by magnetic-field compression, but instead reflects changes in the magnetic-field structure. The exceptionally high fractional polarization then implies a dominance of the ordered magnetic-field component over the turbulent component in the island region. Large-scale tidal stretching and shear can naturally increase the ordering of magnetic fields without substantially changing the total field strength, making tidal interaction a plausible mechanism for producing such a highly polarized synchrotron ridge.

Similar highly polarized magnetic ridges have been observed in several environmentally disturbed nearby galaxies, particularly in the Virgo Cluster \citep[e.g.][]{hummel1995,chyzy2007,chyzy2008,vollmer2007,vollmer2008,choi2026}. These studies demonstrate that environmental interactions can strongly modify the magnetic-field structure and produce extended polarized synchrotron ridges, often without a corresponding increase in total magnetic-field strength. In particular, NGC~4522 illustrates that environmental effects can simultaneously reorganize the magnetic field and affect the spatial distribution and spectral properties of the radio emission \citep{choi2026}. While ram pressure is the dominant mechanism in many Virgo cluster systems, these examples more generally highlight the role of large-scale environmental forces in ordering magnetic fields.

One alternative possibility is that the island represents a fossil synchrotron structure left by a past episode of local star formation. The CRE radiative lifetimes estimated in Sect.~\ref{subsec:smooth_exp} are only $\sim 23$~Myr at 943~MHz and $\sim 10$~Myr at 5~GHz. Therefore, such an episode would need to have occurred within the last $\sim 10$--$30$~Myr to maintain the observed GHz synchrotron emission, unless re-acceleration or compression is present. A recent local episode of this kind would likely leave residual star-formation tracers, especially in the UV. H$\alpha$ emission traces the ionizing photons from short-lived massive stars and is therefore sensitive to very recent star formation on timescales of $\lesssim 5$--$10$~Myr. In contrast, the UV continuum remains sensitive to young stellar populations over longer timescales, of order $\sim 100$~Myr, with NUV generally probing slightly older populations than FUV~\citep{kennicutt1998,kennicutt2012,calzetti2013}. The absence of clear H$\alpha$, infrared, FUV, and NUV counterparts in the island region (Fig.~\ref{fig:NGC2442_RGB_map}) therefore makes a purely fossil origin less likely, although it cannot be completely ruled out.

Future observations will be essential to further constrain the origin of the island and the peculiar morphology of NGC~2442. High-resolution {\sc H\,i} imaging and kinematic modeling are needed to determine whether the island is associated with gas compression, displacement, or stripping, and to distinguish the relative roles of tidal interaction and ram pressure. Deeper H$\alpha$/IFU and X-ray observations would help identify weak shocked or extraplanar gas associated with environmental interaction, while CO observations could reveal whether compressed molecular gas is present in the vicinity of the island.

\section{Conclusions}
\label{sec:conclu}

We have investigated the propagation of CREs in the nearby face-on spiral galaxy NGC~2442 using new and archival multi-frequency RC data, together with optical and infrared tracers of the star-forming environment. Our main conclusions are as follows:

\begin{enumerate}
    
    \item The integrated RC spectrum of NGC~2442 is steeper than that of most nearby spiral galaxies, with spectral indices of $\alpha=-0.96\pm0.04$ for the total emission and $\alpha_{\rm nt}=-1.21\pm0.04$ for the synchrotron emission over 408~MHz--5~GHz. The full spectrum is better described by a broken power law with a break near 1~GHz, suggesting substantial radiative losses in the CRE population.

    \item The thermal emission derived from the H$\alpha$ and 24~$\mu$m data shows that the spiral arms and the central region have enhanced thermal radio emission, while the thermal fraction remains relatively low along the spiral arms. This implies that the RC emission in the disk is largely dominated by synchrotron radiation.

    \item From the spatially resolved synchrotron spectral index map and the equipartition assumption, we derived a mean total magnetic-field strength of $10.8\,\mu{\rm G}$. Within the actively star-forming spiral-arm region used for the RC--SFR analysis, the average field strength is $12.4\,\mu{\rm G}$.

    \item A localized steep-spectrum synchrotron ``island'' is identified in the southeastern region of NGC~2442, with an average synchrotron spectral index of $\alpha_{\rm nt}\approx -1.09$ and no H$\alpha$, infrared, FUV or NUV counterpart. This suggests that the emitting CREs are unlikely to be injected in situ. Our 2D CRPropa simulations show that the island can be explained naturally by anisotropic diffusion along ordered magnetic fields.

\end{enumerate}

Overall, NGC~2442 provides a clear example of how ordered magnetic fields and environmental disturbances can jointly regulate CRE propagation in spiral galaxies. In particular, it shows that in disturbed systems, anisotropic diffusion along ordered magnetic structures can redistribute CREs over kiloparsec scales and produce synchrotron features that are not directly associated with ongoing star formation.

\begin{acknowledgements}
We would like to thank the critical comments by the reviewer which have significantly improved this paper. This research has been supported by the National SKA Program of China (2022SKA0120101, 2022SKA0120103). J.T.L. acknowledges financial support from the National Natural Science Foundation of China (NSFC) through the grants 12321003 and 12273111, the science research grants from the China Manned Space Program with grant Nos. CMS-CSST-2025-A04 and CMS-CSST-2025-A10, and Jiangsu Innovation and Entrepreneurship Talent Team Program through the grant JSSCTD202436. AS is supported by the Australian Research Council through the Discovery Early Career Researcher Award (DECRA) Fellowship (project DE250100003) funded by the Australian Government. This scientific work uses data obtained from Inyarrimanha Ilgari Bundara / the Murchison Radio-astronomy Observatory. We acknowledge the Wajarri Yamaji People as the Traditional Owners and native title holders of the Observatory site. The Australian SKA Pathfinder is part of the Australia Telescope National Facility which is managed by CSIRO. Operation of ASKAP is funded by the Australian Government with support from the National Collaborative Research Infrastructure Strategy. Establishment of ASKAP, the Murchison Radio-astronomy Observatory and the Pawsey Supercomputing Centre are initiatives of the Australian Government, with support from the Government of Western Australia and the Science and Industry Endowment Fund. 

\end{acknowledgements}
\bibliographystyle{aa}
\bibliography{ngc2442}

\begin{appendix}
\section{Magnetic-field model of NGC~2442}
\label{sec:B_model}

We model the magnetic field in the disk plane using a simplified
two-dimensional parametric description. The model is intended to
provide an illustrative field geometry for the CRE transport simulations, rather than a unique reconstruction of the true magnetic-field structure of NGC~2442. We therefore separate the magnetic-field strength and direction as
\begin{equation}
    \mathbf{B}(x,y) = B_{\rm eq}(x,y)\,\hat{\mathbf{b}}(x,y),
\end{equation}
where $B_{\rm eq}(x,y)$ is used only to prescribe the large-scale spatial variation of the total magnetic-field strength, while $\hat{\mathbf{b}}(x,y)$ represents an idealized ordered-field direction. The field strength is taken from the two-dimensional equipartition magnetic-field map shown in Fig.~\ref{fig:B_distri} and is interpolated to arbitrary positions in the $x$--$y$ plane. This choice should be regarded as an approximation to the large-scale distribution of the total field-strength. Since $B_{\rm eq}$ is derived from the total synchrotron intensity, it cannot by itself distinguish between ordered and turbulent magnetic-field components. 

\begin{figure}[!htbp]	
    \centering    \includegraphics[width=0.98\columnwidth]{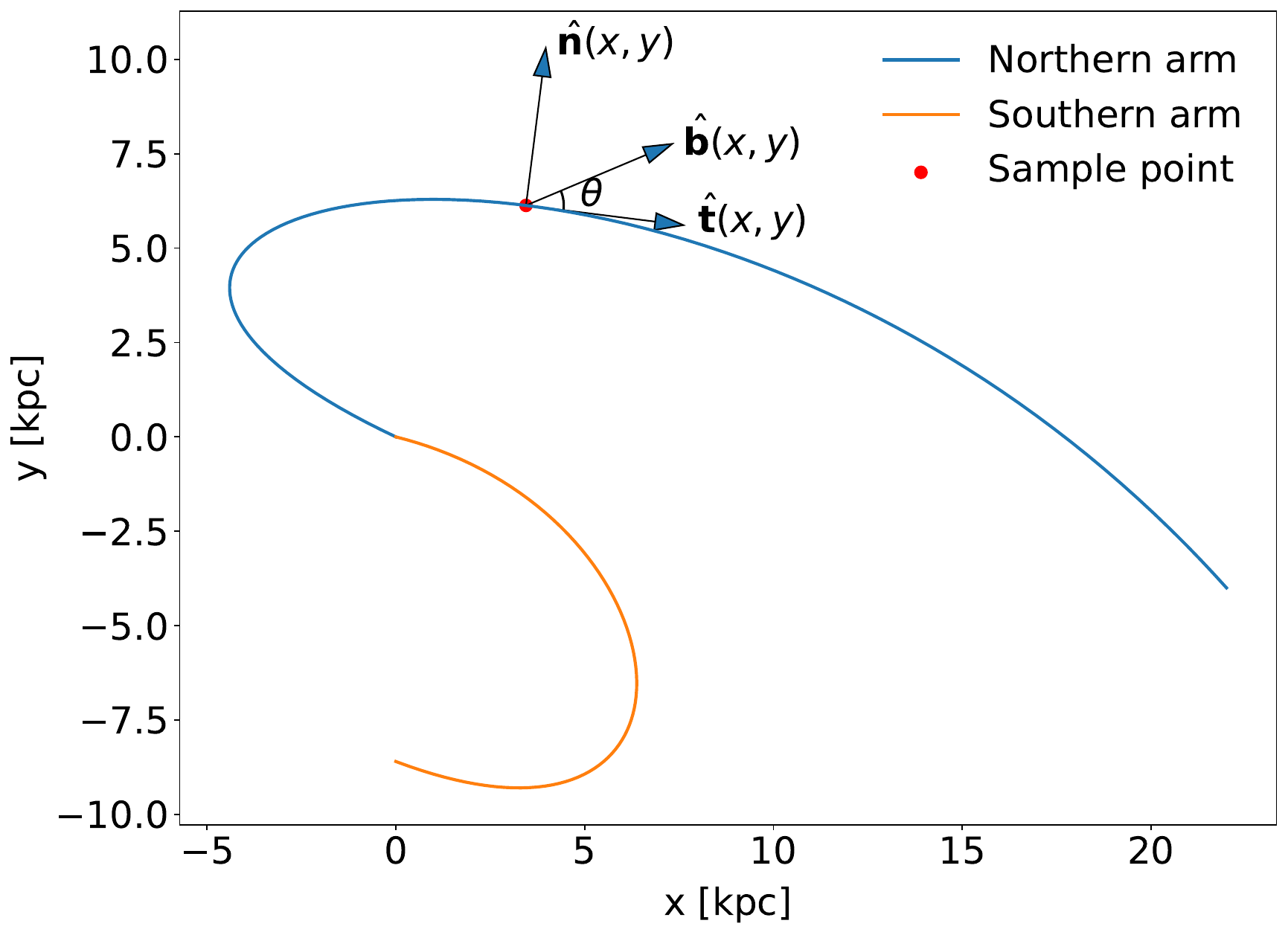}
    \caption{Schematic illustration of the offset angle $\theta$ in the disk plane. At the sample point on the northern spiral arm, $\hat{\mathbf{t}}(x,y)$ denotes the local tangent direction of the arm, $\hat{\mathbf{n}}(x,y)$ is the corresponding in-plane normal direction, and $\hat{\mathbf{b}}(x,y)$ is the magnetic-field direction obtained by rotating $\hat{\mathbf{t}}(x,y)$ toward $\hat{\mathbf{n}}(x,y)$ by an angle $\theta$.} 
\label{fig:ngc2442_theta_schematic}
\end{figure}

The magnetic-field direction is constructed from a smooth, distance-weighted combination of the tangent vectors of the two spiral arms. Let $\hat{\mathbf{t}}_N(x,y)$ and $\hat{\mathbf{t}}_S(x,y)$ denote the local unit tangent vectors of the northern and southern spiral arms, respectively. At a given position $(x,y)$, we first determine the shortest distances from that point to the northern and southern spiral arms, denoted by $d_N(x,y)$ and $d_S(x,y)$. The relative influence of the two arms is then described by Gaussian distance weights,
\begin{equation}
w_N(x,y)=\exp\left[-\frac{d_N^2(x,y)}{2\sigma_{\rm arm}^2}\right],
\qquad
w_S(x,y)=\exp\left[-\frac{d_S^2(x,y)}{2\sigma_{\rm arm}^2}\right],
\end{equation}
where $\sigma_{\rm arm}$ is a smoothing scale that controls how rapidly the influence of each spiral arm decreases with distance from the arm ridge. In this formulation, locations closer to a given spiral arm are more strongly influenced by its local tangent direction, while positions farther away receive a progressively smaller contribution. The local in-plane tangent direction is then written as
\begin{equation}
\hat{\mathbf{t}}(x,y)=
\frac{w_N(x,y)\,\hat{\mathbf{t}}_N(x,y)+w_S(x,y)\,\hat{\mathbf{t}}_S(x,y)}
{\left|w_N(x,y)\,\hat{\mathbf{t}}_N(x,y)+w_S(x,y)\,\hat{\mathbf{t}}_S(x,y)\right|}.
\end{equation}
The corresponding in-plane normal unit vector is $\hat{\mathbf{n}}(x,y)=(-t_y,t_x)$, where $\hat{\mathbf{t}}(x,y)=(t_x,t_y)$. Allowing for an offset angle $\theta$ between the magnetic field and the local spiral-arm tangent (see Fig.~\ref{fig:ngc2442_theta_schematic}), the final magnetic-field direction in the disk plane is
\begin{equation}
\hat{\mathbf{b}}(x,y)=
\cos\theta\,\hat{\mathbf{t}}(x,y)
+
\sin\theta\,\hat{\mathbf{n}}(x,y).
\end{equation}

Adopting $\theta = 10\degr$, the resulting magnetic-field orientation in the disk plane is nearly parallel to the local spiral-arm structure. The resulting two-dimensional magnetic-field directions and strengths are shown in Fig.~\ref{fig:ngc2442_2D_Bfield}.

\section{Energy loss during CRE propagation}
The propagation of CREs was modeled with CRPropa using anisotropic diffusion in the prescribed two-dimensional magnetic field. Synchrotron losses and IC losses on the CMB were treated directly within CRPropa during each propagation step. In addition, bremsstrahlung, ionization, and IC losses associated with the stellar and infrared radiation fields were included through an operator-splitting scheme applied after each outer propagation interval $\Delta t_{\rm step}$.

The IC loss timescale can be written as~\citep{schleicher2013}
\begin{equation}
\left(\frac{\tau_{\rm ic}}{\rm yr}\right)=
5.7 \times 10^{7}
\left(\frac{\nu_{\rm c}}{\rm GHz}\right)^{-1/2}
\left(\frac{B}{\mu{\rm G}}\right)^{1/2}
\left(\frac{U_{\rm rad}}
{10^{-12}\ {\rm erg\ cm^{-3}}}\right)^{-1},
\end{equation}
where $U_{\rm rad}$ is the total radiation energy density, as discussed in Sect.~\ref{subsec:smooth_exp}.

Additional energy losses due to ionization were included with a characteristic timescale~\citep{schleicher2013}
\begin{equation}
\left(\frac{\tau_{\rm ion}}{\rm yr}\right)=
4.1 \times 10^{9}
\left(\frac{E}{\rm GeV}\right)
\left(\frac{n_{\rm ISM}}{\rm cm^{-3}}\right)^{-1}
\left(3\ln\frac{E}{\rm GeV}+42.5\right)^{-1},
\end{equation}
where $n_{\rm ISM}$ is the number density of the interstellar medium. The CRE energy at the observing frequency $\nu$ can be calculated from~\citep{beck2015}
\begin{equation}
    E(\rm GeV)\approx\sqrt{\frac{\nu\,(\rm MHz)}{16\,\mathrm{MHz} \times B_{\perp}\,(\rm \mu G)}},
\label{eq:v_to_E}
\end{equation}
where $B_{\perp}$ is the perpendicular magnetic field strength, which can be approximated as $B_{\perp}=\sqrt{2/3}\,B_0$ for an isotropic turbulent magnetic field, where $B_0$ is the total magnetic field strength in the disc plane. Bremsstrahlung losses were included with a timescale~\citep{schleicher2013}
\begin{equation}
\left(\frac{\tau_{\rm brems}}{\rm yr}\right)=
8.6 \times 10^{7}
\left(\frac{n_{\rm ISM}}{\rm cm^{-3}}\right)^{-1}.
\end{equation}

In the numerical implementation, after each outer propagation step $\Delta t_{\rm step}$, the particle energy was updated according to
\begin{equation}
E_{n+1}=E_{n+1}^{\rm CRPropa}-\Delta E_{\rm ion}-
\Delta E_{\rm brems}-\Delta E_{\rm ic,\,star+TIR},
\end{equation}
where the individual corrections were evaluated from
\begin{equation}
\Delta E_{j}=E\,\frac{\Delta t_{\rm step}}{\tau_{j}}
\qquad
(j={\rm ion,\ brems,\ IC}),
\end{equation}
using the corresponding loss timescales defined above. Here $E_{n+1}^{\rm CRPropa}$ denotes the particle energy after the CRPropa propagation step including diffusion, synchrotron losses, and IC losses on the CMB.

\end{appendix}
\end{document}